\documentclass[iop]{emulateapj}             
\usepackage{natbib}
\shortauthors{Radford \& Peterson}
\shorttitle{Submillimeter Atmospheric Transparency}

\slugcomment{Submitted to PASP: 2015 November 20; Accepted 2016 February 17; update 2016 February 26} 

\begin{document}

\title{\null\vskip1in
Submillimeter Atmospheric Transparency\\ 
at Maunakea, at the South Pole, and 
at Chajnantor
}

\author{ 
Simon J. E. Radford
}\affil{ 
Caltech Submillimeter Observatory,
California Institute of Technology,\\ 
111 Nowelo Street, Hilo, HI 96720\\
and National Radio Astronomy Observatory, Tucson, AZ}
\email{sradford@caltech.edu}

\author{
Jeffery B. Peterson
}\affil{Department of Physics,
Carnegie Mellon University\\
5000 Forbes Avenue, Pittsburgh, PA  15213\\
}
\email{jbp@cmu.edu}

\begin{abstract}

For a systematic assessment of submillimeter observing conditions at different sites,
we constructed tipping radiometers to measure
the broad band atmospheric transparency in the window 
around 350\,$\mu$m wavelength. 
The tippers were deployed  
on Maunakea, Hawaii,
at the South Pole, and
in the vicinity of Cerro Chajnantor in northern Chile.
Identical instruments permit direct comparison of these sites.
Observing conditions at the South Pole and in the Chajnantor area are better
than on Maunakea.
Simultaneous measurements with two tippers demonstrate
conditions at the summit of Cerro Chajnantor are significantly better
than on the Chajnantor plateau.

\end{abstract}


\newpage

\section{Introduction}

At submillimeter wavelengths,
observations from the ground are practical 
at only a handful of exceptional sites
because Earth's atmosphere is only partially transparent.
Pressure broadened transitions of atmospheric molecules,
water vapor in particular, delimit discrete spectral windows.
Within these windows, the line wings both absorb incoming radiation,
attenuating astronomical signals,
and emit thermally, contributing to the background. 

Because water vapor is the primary cause of atmospheric opacity, the
best sites for submillimeter wavelength astronomy have exceptionally dry air. 
With a typical (exponential) scale height of 1--2\,km 
\citep{holdaway:1996, turner:2001},
water is concentrated in the lower troposphere. Hence
extremely dry air can be found above high altitude sites, 
particularly in subtropical desert regions.
In addition, at the low temperatures encountered on the Antarctic interior plateau 
even saturated air has a very small water vapor content.

For a systematic assessment of observing conditions at different sites,
we constructed four tipping radiometers (tippers) to measure
the broad band atmospheric transparency in the 350\,$\mu$m window.
The instruments were deployed  
on Maunakea, at the South Pole,
and in the vicinity of Cerro Chajnantor in northern Chile (\S 3).
This paper describes the instrument design and construction, 
the deployments, and the measurements;
discusses the site characteristics;
and compares the results with data from other instruments 
and with model predictions.
Previous accounts of these measurements  
presented preliminary findings
\citep{radford:2002,peterson:2003,radford:2008,radford:2011}.

\section{Instrument}

\begin{figure}
\begin{center}
\includegraphics[width=0.8\columnwidth]{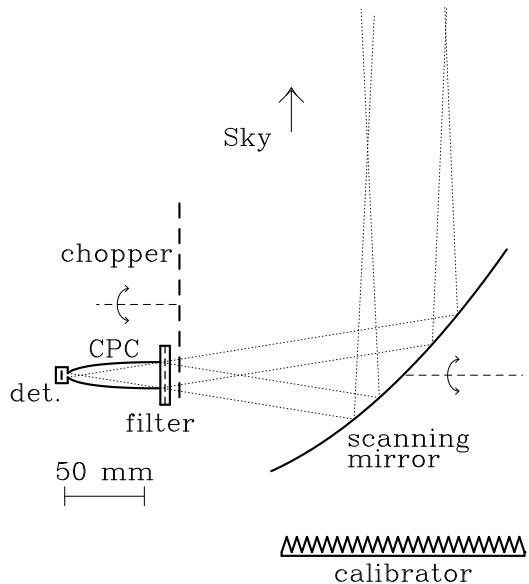} 
\caption[layout]{\label{fig:layout}
Optical layout of submillimeter tipper (to scale).
} 
\end{center}
\end{figure}

The tippers 
(Fig.~\ref{fig:layout}) 
are built around an ambient temperature, 
deuterated L-alanine doped triglycine sulfate 
(DLATGS) pyroelectric detector
with a sensitivity about 4\% of the thermodynamic limit
\citep{putley:1980}.
A light tight box holds the detector
at the exit aperture of a compound parabolic 
concentrator (CPC; \citealp{welford:1989}).
The detector signal is 
buffered by a JFET integrated into the detector package, 
amplified, and digitized at 200\,samples\,s$^{-1}$.

\begin{figure}
\begin{center}
\includegraphics[width=0.8\columnwidth]{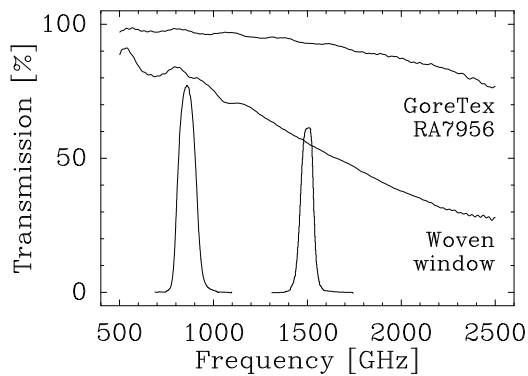} 
\caption[window]{\label{fig:window}
Submillimeter transmission of 0.25\,mm thick woven fabric used
as the tipper's weather cover window compared with 
unwoven GoreTex RA\,7956 sheet
and with passbands of 350\,$\mu$m and 200$\,mu$m filters.
Measurements of fabric made with an spectrometer
in the SAO submillimeter receiver lab
(S. Paine 1998, private communication). 
}
\end{center}
\end{figure}

A resonant metal mesh bandpass filter 
(QMC Instruments) 
mounted at the CPC entrance aperture 
defines the tipper's spectral response
and a comounted 
plastic lowpass filter rejects unwanted mid-IR radiation.
The IR rejection was confirmed by measurements with a laboratory spectrometer.
The filter has a 103\,GHz (FWHM) passband centered at 850\,GHz
(Fig.~\ref{fig:window})
that matches the 350\,$\mu$m atmospheric window.
This passband is the same as 
the UKT14 bolometer \citep{duncan:1990} 
and SHARC \citep{hunter:1996, hunter:1997, benford:1998}, 
is wider than SCUBA \citep{holland:1999},
but is narrower than  SharcII \citep{dowell:2003}.
On three tippers,   
the filters are fixed directly to the CPC aperture.
The fourth instrument, one of the two deployed in Chile,
has a filter wheel that accommodates an additional filter
with an 74\,GHz (FWHM) passband centered at 1500\,GHz
that matches the 200\,$\mu$m atmospheric window.

Because the detector is intrinsically differential, a
rotating blade (chopper) modulates the incident radiation at the CPC entrance. 
The blade has three vanes to avoid even harmonics of its DC drive motor.  
A servo amplifier maintains the chopper at 0.75\,Hz, 
where the detector is most sensitive.  

The CPC entrance aperture lies in the focal plane of a 90\arcdeg\
offset parabolic scanning mirror (Fig.~\ref{fig:layout}). 
Together the CPC and the offset mirror
define the tipper's 6\arcdeg\ beam (FWHM).
Laboratory measurements with a transmitter confirmed 
the beam shape.
The scanning mirror rotates about the CPC axis so the detector views 
the sky from the zenith to the horizon or either of 
two calibrators.
To preclude spillover onto the ground, etc., 
the scanning mirror and the calibrators are substantially underilluminated.

\begin{deluxetable}{p{0.1\columnwidth}p{0.05\columnwidth}cccc} 
\tablewidth{0pt}
\tablecaption{\label{tab:sites}
Site Characteristics} 
\tablehead{
&& \colhead{Mauna-}& 
 \colhead{South}& 
 \colhead{Chaj.}& 
 \colhead{Cerro}\\
&& \colhead{kea}& 
 \colhead{Pole}& 
 \colhead{plateau}& 
 \colhead{Chaj.}
} 
\startdata
Lat. &N &   
    $\phn19\arcdeg\, 48'$&
    $-90\arcdeg$&
    $-23\arcdeg\, \phn1'$&  
    $-22\arcdeg\, 59'$\\  
Long. &W & 
    $155\arcdeg\, 27'$&
    \nodata&
    $\phantom{-}67\arcdeg\, 45'$&
    $\phantom{-}66\arcdeg\, 44'$\\
Noon &UT & 
    $\phantom{-}22^{\rm h}$\, 22$^{\rm m}$&
    \nodata&
    $\phantom{-0}16^{\rm h}$\, 31$^{\rm m}$ &
    $\phantom{-0}16^{\rm h}$\, 31$^{\rm m}$ \\
Alt. &[m] & 4100& 2835& 5060& 5612\\ 
$T_{\rm median}$ &[\arcdeg C] 
& 
    $2$\,\tablenotemark{a}&
    $-50$\,\tablenotemark{c}& 
    $-2$\,\tablenotemark{d}&
    $-6$\,\tablenotemark{e}\\
WS${}_{\rm med.}$\tablenotemark{*} &[m\,s$^{-1}$]
&
    7\,\tablenotemark{b}&
    $\phantom{-0}6$\,\tablenotemark{c}&
    $\phantom{-}6$\,\tablenotemark{d}&
    $\phantom{-}4$\,\tablenotemark{e}
\enddata
\strut
\tablenotetext{*}{wind speed, median}
\tablenotetext{a}{\citealt{businger:2002}}
\tablenotetext{b}{\citealt{erasmus:1986}}
\tablenotetext{c}{\citealt{king:1997}}
\tablenotetext{d}{\citealt{radford:1998}}
\tablenotetext{e}{\citealt{radford:2008}}
\end{deluxetable}

Each calibrator is made from a 12\,mm thick slab of epoxy 
loaded with 5\,$\mu$m iron beads (Emerson \& Cuming Eccosorb MF-110).
This material is the machinable form of the 
castable CR-110 used for several previous
experiments, including the
COBE mission. At 300\,K and 350\,$\mu$m, the specular reflectance is
about 10\% and the index of refraction is 
about 1.9 \citep{peterson:1984, hemmati:1985}. 
In the tipper calibrators, 
concentric triangular grooves are machined in the face
of the epoxy slab.
The (full) groove angle is 37\arcdeg, so
rays incident on the calibrator 
within 10\arcdeg\ of normal are reflected 
five or more times before reemerging.
Moreover, the Brewster angle is 62\arcdeg\ for this material,
so one polarization is almost completely absorbed. 
We estimate the overall specular reflectance of 
the calibrators is $\le 10^{-5}$.
On the sides and rear, the grooved epoxy absorber is attached 
to 6.4\,mm thick aluminum plates and
an electric heater is bonded to the other side
of the rear plate.
An analog servo circuit regulates the heater.
The calibrator assembly's rear and side surfaces 
are insulated with 19\,mm of polystyrene foam. 
The calibrator's front surface is insulated with  
two thin 
(0.19\,mm) expanded PTFE (Zitex ZA-105) membranes
suspended
3.2\,mm and 6.4\,mm above the calibrator surface. 
This material has very low 
attenuation throughout the submillimeter
\citep{benford:1999}. 
One calibrator (hot) is heated to $\approx 330$\,K while the other (warm) is 
allowed to equilibrate with the tipper's internal temperature.
On Maunakea and in Chile, the tippers' internal 
temperatures equilibrate about 20--25\,K above the
surroundings and follow the diurnal and seasonal cycles.
At the South
Pole, the tipper's internal temperature is maintained
at $(284 \pm 1)$\,K by thermostatically controlled electric heaters.

The tipper is mounted in a weatherproof, insulated
enclosure 
and views the sky through a 
window of 0.25\,mm thick
woven fabric similar to GoreTex. 
Across the 350\,$\mu$m filter passband, the window transmission 
is $(80 \pm 3)$\% 
(Fig.~\ref{fig:window}).
Across the 200\,$\mu$m filter passband, the window transmission is about 55\%.

A computer operates the tipper,
processes the data, and records the results.
In particular, the synchronous demodulation 
and integration of the detector signal 
is done by software.

\section{Deployments}
 
The tippers were deployed to four premier locations for submillimeter 
astronomy
(Table~\ref{tab:sites}).

Maunakea is the highest isolated marine mountain in the world.
A stable inversion layer often traps moisture
below the summit, especially at night.
Excellent observing conditions
have led to the development of major astronomical facilities.
Maunakea has become a standard of comparison for observatory sites  
\citep{morrison:1973, erasmus:1986, businger:2002}.
In 1997 December, a submillimeter tipper was installed 
pointing northwest 
on the roof of an outbuilding
at the Caltech Submillimeter Observatory (CSO)
adjacent to an existing 225\,GHz tipper. 
In 2016, the instrument will be relocated to the roof of the 
Submillimeter Array (SMA) hangar, where it will continue to point northwest.

At the South Pole
the atmospheric water vapor content is very small 
because of the extremely low temperature
even though the surface air 
is almost saturated
\citep{gettelman:2006}. 
Several telescopes have been installed there,
primarily for millimeter and submillimeter wavelength astronomy and cosmology 
\citep{burton:2012}.
In 1998 January, a submillimeter tipper was installed on the roof of the 
AST/RO building where it pointed approximately along $135^\circ$ west longitude.
In 2006 January, the instrument was relocated to the roof of the
Dark Sector Lab (DSL) where it points along $10^\circ$ east longitude.
An additional, fifth, instrument, with a different detector and other modifications, 
was deployed to Dome C, Antarctica, in 2000--1 and in 2003 with 
a comparison period at the South Pole in 2001--2 
\citep{calisse:2004b,calisse:2004c}.

In northern Chile, the combined effects of a high pressure belt over the 
southeast Pacific, a cold offshore current, and the 
moisture barrier of the Andean cordillera to the east
make the Atacama desert one of the driest places on
Earth.  
The absence of glaciers, even on the highest peak in the region,
Volc\'an Llullaillaco (6740\,m), is unique for these altitudes
\citep{messerli:1993} and attests to the area's aridity.
In recent years, the Chajantor plateau east of the village of 
San Pedro de Atacama
has seen the installation of several radiotelescopes, culminating
in the international Atacama Large 
Millimeter/submillimeter Array (ALMA; \citealp{beasley:2006}). 
In 1997 October, a submillimeter tipper was installed pointing east adjacent 
to an existing 225\,GHz tipper near the (then future) center of ALMA.
In 2000 June, a second instrument was deployed to the same location.
In 2005 May, the instruments were relocated about 1\,km west to 
the site of the CBI \citep{padin:2001} and successor experiments,
where they pointed south.
In 2011 November, one instrument was relocated to the 
APEX telescope \citep{gusten:2006}, 
about 4\,km north of ALMA, again pointing south.

Several peaks rise above the Chajnantor plateau.
Radiosondes launched from the plateau 
determined the typical (exponential) water vapor scale height is
about 1.1\,km
and also revealed frequent inversion layers, 
especially at night,
that trap moisture below the tops of these peaks
\citep{giovanelli:2001}.
Subsequent measurements
in the submillimeter  
\citep{blundell:2002,marrone:2004,marrone:2005,bustos:2014}
and in the near infrared \citep{konishi:2015} 
have buttressed this result, 
finding exceptional observing conditions on these peaks. 
In 2006 May, one tipper was installed pointing south
near the summit of one peak, Cerro Chajnantor,
550\,m above and 8\,km northeast of ALMA.

\section {Measurements}

The tipper determines the submillimeter atmospheric transparency 
by the standard technique of
measuring the sky brightness at several zenith angles
\citep{dicke:1946}. 
During each tip, which takes about 13 minutes,
the demodulated detector signal is integrated for 64 s at 
each of
eleven positions of the scanning mirror: pointing at 
the hot calibrator; at the warm calibrator; 
at airmasses of 1, 1.5, 2, 2.5, 3, 3.5, and 4 
(zenith angles of 0$\arcdeg$, 48$\arcdeg$, 
60$\arcdeg$, 66$\arcdeg$, 71$\arcdeg$, 73$\arcdeg$, 
\& 75.5$\arcdeg$);
and at the hot and warm calibrators again. 
The tipper makes single sided tips from
the zenith down towards one horizon.
During installation 
the tippers were leveled within $\pm 2.5^\circ$.

\subsection {Calibration}

At the start and again at the end of each tip cycle, 
the two calibrators are observed to determine
the detector responsivity, 
\begin{equation} 
  R = (V_{\rm hot} - V_{\rm warm}) / (T_{\rm hot} - T_{\rm warm}) \ , 
\end{equation}
and the effective chopper temperature, 
\begin{equation} 
  T_{\rm chop} = T_{\rm hot} - V_{\rm hot} / R = T_{\rm warm} - V_{\rm warm} / R \ . 
\end{equation}
Here $V_{\rm hot}$ and
$V_{\rm warm}$ are the demodulated detector signals for the hot and warm calibrators
and $T_{\rm hot}$ and $T_{\rm warm}$ are their physical temperatures
recorded during each calibration measurement. 
The effective chopper temperature, $T_{\rm chop}$, is the temperature
seen by the detector when the chopper blade blocks the 
CPC entrance aperture.
For each tip cycle, the average of 
the 
two calibration measurements is used to calibrate the sky measurements. 

To check the calibrator performance, an absorber immersed in liquid nitrogen
was substituted for the hot calibrator.
The detector responsivity determined with the normal calibration and 
with the LN$_2$ target agreed within 10\%.

Although the detector's responsivity is somewhat temperature dependent,
the tipper's large thermal inertia precludes rapid changes
so the transparency measurements are unaffected.
Moreover, whenever there was an excessive difference between
the detector responsivity measured at the
start and at the end of each tip cycle, the observation was discarded.

\subsection {Window}

The tipper views the sky through a fabric window but 
the calibrators are inside this window.
Ignoring the window
would cause an underestimate
of the atmospheric transparency
\citep{calisse:2004a}.
We model the window 
as an absorbing screen at the ambient exterior air temperature, 
$T_{\rm ext}$.
Then the brightnesses measured during a tip,
\begin{eqnarray}
  T_{\rm meas}(A) &=& \Delta V_{\rm sky}(A) / R +  T_{\rm chop}  \nonumber \\
                  &=& \theta_{\rm win}T_{\rm sky}(A) + T_{\rm win}  \ ,
\end{eqnarray}
where
$\Delta V_{\rm sky}$ is the demodulated detector signal measured at airmass $A$,
 $T_{\rm sky}$ is the sky brightness,
$ \theta_{\rm win}$ is the window transmission,
 and
$ T_{\rm win} = T_{\rm ext} ( 1 - \theta_{\rm win}) $ is 
the window brightness.

To check the window correction, we temporarily placed a second layer of
window material over the tipper at the South Pole
while conditions were both good, $\tau \approx 0.7$,
and stable. 
In five trials of this experiment, the apparent zenith optical depth 
increased by $0.18 \pm 0.07$, consistent
with the calculated correction \citep{peterson:2003}.

A digital sensor (DS 1820) measured the exterior air 
temperature, $T_{\rm ext}$, at the start
and end of each tip cycle; the average was used 
for the window correction.
This sensor  has an absolute minimum 
reading of $-55\,\arcdeg$C (218\,K). 
During the Antarctic winter the actual air temperature can 
fall below this.
Hence the window temperature might be overestimated, leading to an
overestimate of the atmospheric transparency. 
In practice, however, this is a small effect,
less than other measurement uncertainties.
If the exterior temperature were actually $-80\,\arcdeg$C when the sensor measures 
$-55\,\arcdeg$C, the window brightness temperature
would be 3 K colder, and the apparent 
optical depth would be reduced by less than 3\%. 

\subsection {Atmospheric Transparency}

\begin{figure} 
\begin{center}
\includegraphics[
  width=\columnwidth
  ]{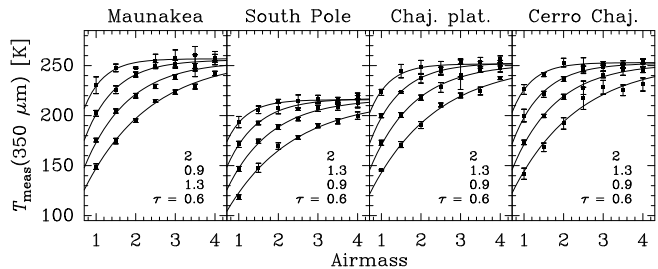} 
\caption[tips]{\label{fig:tips}
Representative 350\,$\mu$m sky brightness tips 
selected to show the range of conditions encountered
in the measurements.
They are arranged ({\it bottom to top\/})
in order of increasing optical depth 
with lines illustrating the best fit two parameter models
(Eq.~\ref{eq:tip}). 
} 
\end{center}
\end{figure}

The zenith optical depth, $\tau$, is determined by fitting
the measured sky brightnesses to a two parameter model 
of a plane parallel uniform atmosphere, 
\begin{eqnarray}
  T_{\rm meas}(A) - T_{\rm win}  &=&  \theta_{\rm win} T_{\rm sky}(A) \nonumber \\
    &=&  \theta_{\rm win} T_{\rm atm} (1-e^{-\tau A}) \ , \label{eq:tip}
\end{eqnarray}
where $T_{\rm atm}$ is the effective atmospheric brightness
(Fig.~\ref{fig:tips}).
Typically, the uncertainty in the individual sky brightness measurements 
is 5\,K or less, independent of airmass.

Because the brightness curve is non linear, the dynamic range is
limited and the results are heteroscedastic:
the uncertainty in the optical depth depends on the magnitude of the optical depth. 
The fitting procedure is robust over the range $0.5 < \tau < 4 $ 
where there is sufficient curvature to separate the
optical depth from the atmospheric brightness temperature.
When $\tau \approx 1$, the typical uncertainty $\sigma(\tau) < 0.1$.
Under poor conditions, when $\tau > 4$, the brightness difference between
the zenith and the horizon vanishes. 
Then only the product $\theta_{\rm win} T_{\rm atm}$ can be determined
and the uncertainty in optical depth diverges.
A lower bound on the uncertainty is roughly
$\sigma(\tau) > \tau^4$.
For this reason, poor conditions with high optical depth cannot
be quantified well and must be simply regarded as poor.
So long as overall statistics take this into account, this is
not a practical drawback for characterizing observing conditions.
(If the optical depth were very low, $\tau \ll 0.5$, the model 
would become linear so
only the product $\theta_{\rm win} T_{\rm atm} \tau$ could be measured.
At 350\,$\mu$m, however,
such conditions are never encountered in practice.)

Moreover, the atmosphere is not uniform so
the measured sky brightness is weighted 
by the vertical profiles of temperature and optical depth.
At large zenith angle (low elevation), the effective
beam termination is lower
in the atmosphere, where the air
is generally warmer, than the termination at the zenith. 
Indeed, this effect is sometimes present in the data: 
near the horizon the sky temperature doesn't approach 
an asymptote, but continues to rise. 
For simplicity and because of the 
limited degrees of freedom in the data, we  
have ignored this effect in the data analysis.

\subsection{Data Edits}

At various times, the tippers suffered outages 
that might affect the data quality.
Measurements were excised, therefore, if they
failed any of several tests
indicating software or hardware malfunction.
The fraction of data discarded depended on local 
circumstances.
Because these tests are based on objective, internal 
properties of the instruments, not on external conditions,
we believe these edits have not biased the data.
Indeed, the statistics of the raw, unedited data are very close to 
the statistics of the edited data.

\subsection{Side by side comparisons\/}

\begin{figure} 
\begin{center}
\includegraphics[clip=true,
  width=0.45\columnwidth]
  {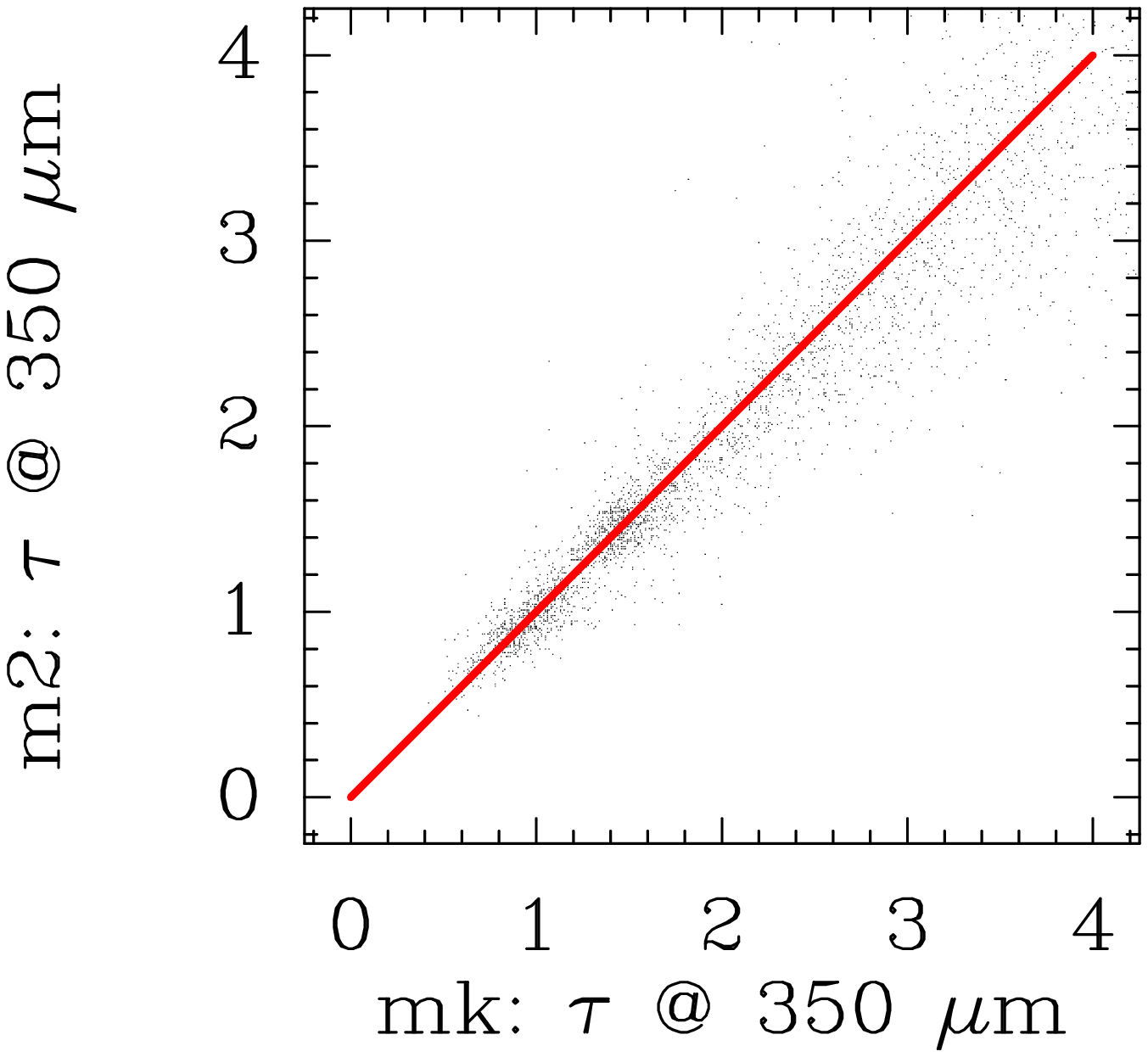}
\quad
\includegraphics[clip=true,
  width=0.45\columnwidth]
  {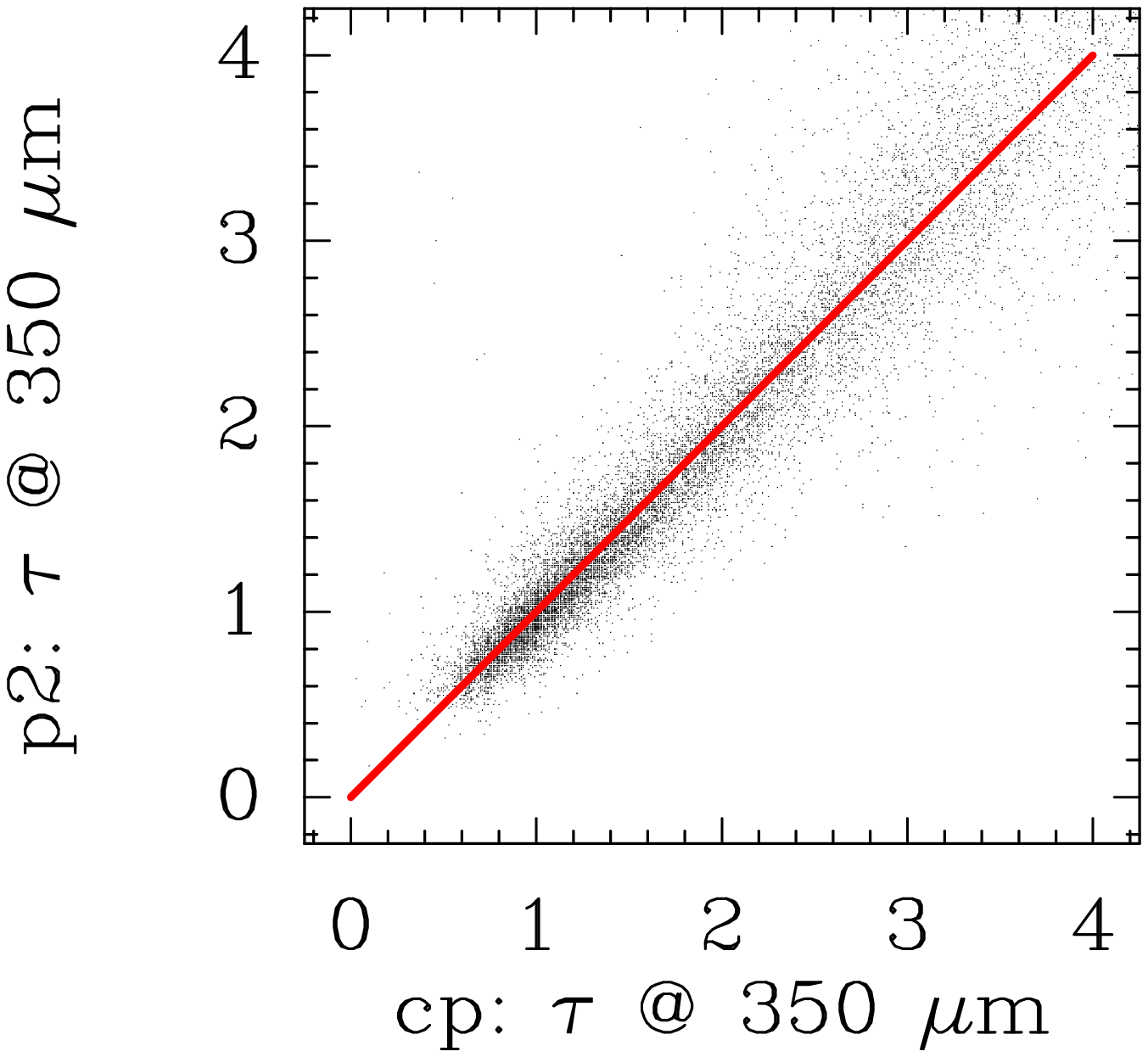}
\caption[cmp-350]{\label{fig:cmp-350}
Correlation between broad band 350\,$\mu$m zenith optical depth measured 
simultaneously with side by side tipper pairs:
({\it left\/}) on Maunakea during 2011 August-October 
($ r = 0.97$) and 
({\it right\/}) on the Chajnantor plateau during parts or all of 31 months in 
2000--1, 2005--6, and 2009
($ r = 0.98$).
} 
\end{center}
\end{figure}

Although the instruments share the same design, components, and construction, 
side by side comparisons are important to confirm 
they perform identically. 
Although it was not possible to test all the tippers together 
under realistic submillimeter observing conditions
prior to deployment, 
comparisons were made on three occasions.
On Maunakea, two
tippers were operated side by side during 2011 August--October.  
On the Chajnantor plateau,
the other two instruments were operated side by side 
during parts or all of 31 months
in 2000--1, 2005--6, and 2009.
This included one week when
the two instruments were separated by 1\,km, one at 
at ALMA and the other at the CBI.
In 2001--2, the South Pole tipper was operated side by side
with the modified tipper used at Dome C
\citep{calisse:2004b,calisse:2004c}.
On all these occasions,
the side by side measurements show excellent agreement
(Fig.~\ref{fig:cmp-350}).
The paired, simultaneous measurements are highly correlated, $r \approx 1$,
and indicate the instruments produce identical results.

\section{Results}

\begin{deluxetable}{lcccc}
\tablewidth{0pt}
\tablecaption{\label{tab:quartiles}
\strut 350\,$\mu$m zenith optical depths 
} 
\tablehead{& Mauna-& South& Chajnantor& Cerro\\
  & kea& Pole& plateau& Chajnantor\\
  \cline{2-5} \\
  \strut
  start& 1997 Dec& 1998 Jan& 1997 Oct& 2006 May\\
  stop&  2016 Feb& 2016 Feb& 2016 Feb& 2013 Jun}
\startdata
75\,\%& 4.2& 1.7& 2.9& 2.0\\
50\,\%& 2.5& 1.3& 1.7& 1.1\\
25\,\%& 1.5& 1.1& 1.1& 0.8
\enddata
\end{deluxetable}

\begin{figure} 
\begin{center}
\includegraphics[width=0.9\columnwidth]{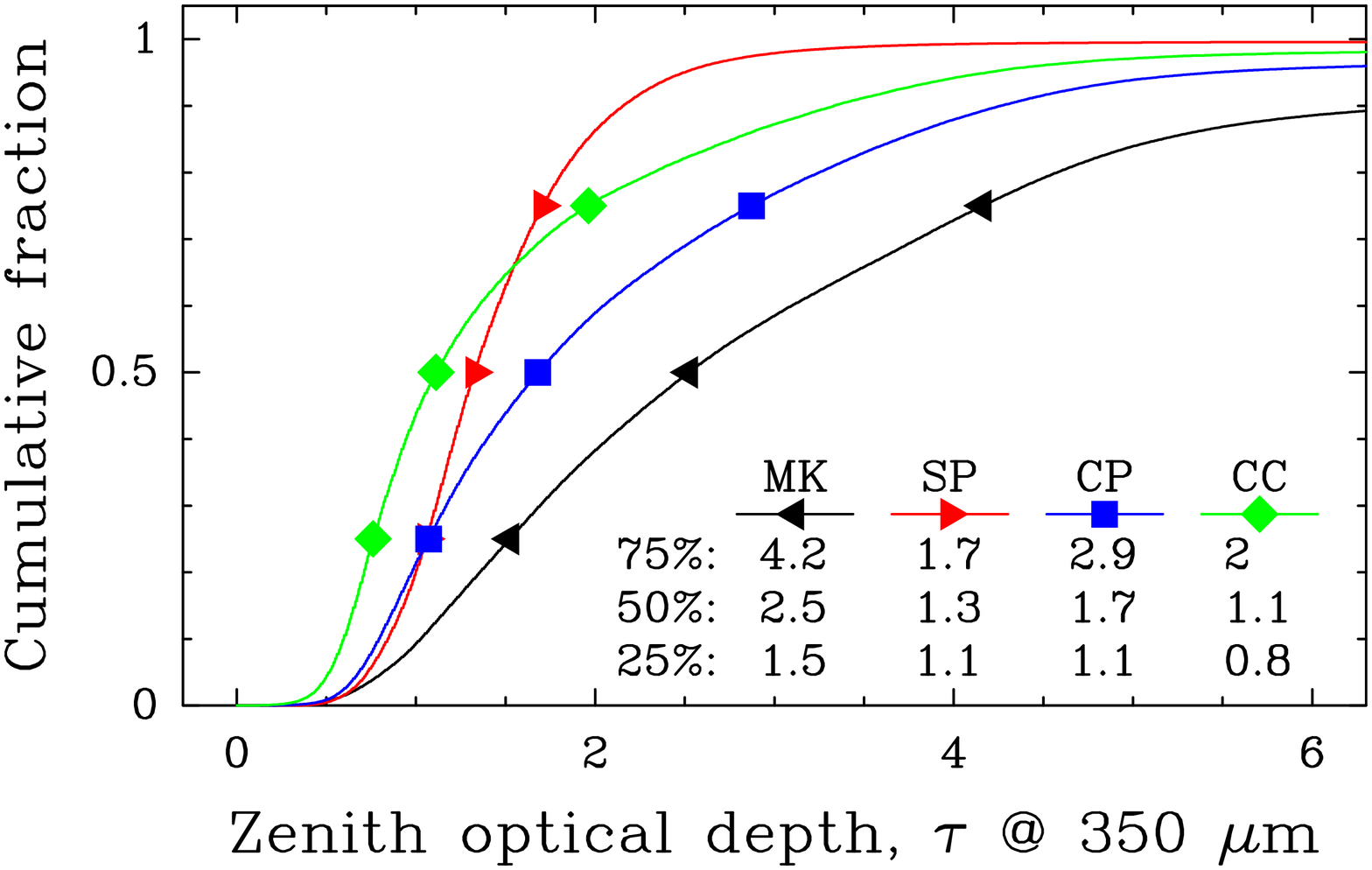} 
\caption[cdf-350]{\label{fig:cdf-350}
Cumulative distributions of broad band
350\,$\mu$m zenith optical depths measured 
on Maunakea (MK), 
at the South Pole (SP), 
on the Chajnantor plateau (CP), and
on Cerro Chajnantor (CC).
}
\end{center}
\end{figure}

The measurements show
all four sites enjoy periods of excellent observing conditions,
$\tau \le 1$
(Table~\ref{tab:quartiles} \& Fig.~\ref{fig:cdf-350}).
Even on these occasions, however, atmospheric absorption
at 350\,$\mu$m is substantial.
As a reminder, when the zenith optical depth $\tau = 1$, 
the zenith transmission is only 37\%.
Submillimeter astronomy remains challenging even at premier sites.

\subsection{Site Conditions}

The best conditions occur more often at the South Pole and in the 
vicinity of Chajnantor than on Maunakea. 
The median optical depth on the Chajnantor plateau is similar 
to the first quartile on Maunakea.
First quartile conditions at the South Pole 
and on the Chajnantor plateau are similar.
The cumulative distribution for the South Pole is remarkably sharp;
the South Pole hardly ever experiences the poor conditions,  $\tau > 3$,
experienced at the other sites during storms.
Conditions on Cerro Chajnantor are significantly better than 
on the the plateau.

At all locations, the minimum measured optical depth is not zero 
but in the range 0.3--0.4. 
Although this measured minimum
may be at least partially an instrumental artifact, 
it does depend on altitude, being
smallest at the highest site, Cerro Chajnantor. 
This suggests the minimum corresponds to absorption 
by dry air, i.\,e., atmospheric components other than water vapor.
This phenomenon has been discussed in detail previously  
\citep{pardo:2001a, pardo:2001b, pardo:2005}.

As perhaps might be expected, 
there is no significant correlation between 
the measurements at widely separated sites, i. e., 
between Maunakea and the South Pole or Chile.
Hence the joint distribution of conditions for simultaneous observations 
is the product of the distributions for the individual sites.
This may be a consideration, for example, for future short wavelength VLBI. 

\begin{figure} 
\begin{center}
\includegraphics[width=0.99\columnwidth]{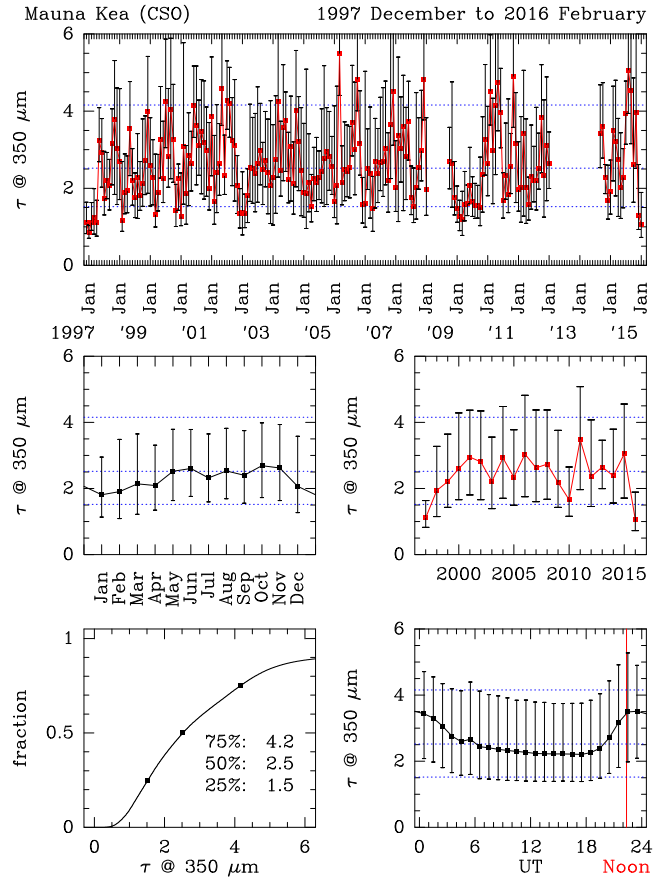}
\caption[mk-0-dat]{\label{fig:mk-0-dat}
Broad band 350\,$\mu$m zenith optical depth measured on Maunakea.
{\it Top:\/} Monthly quartiles (25\%, 50\%, \& 75\%); 
{\it center left:\/} seasonal variation;
{\it center right:\/} yearly quartiles;
{\it bottom left:\/} cumulative distribution;
{\it bottom right:\/} diurnal variation (with mean solar noon indicated).
In each panel, markers indicate median values and the error bars show the 
first and third quartiles (25\% \& 75\%).
Horizontal dotted lines show the overall quartiles. 
}
\end{center}
\end{figure}

\begin{figure} 
\begin{center}
\includegraphics[width=0.99\columnwidth]{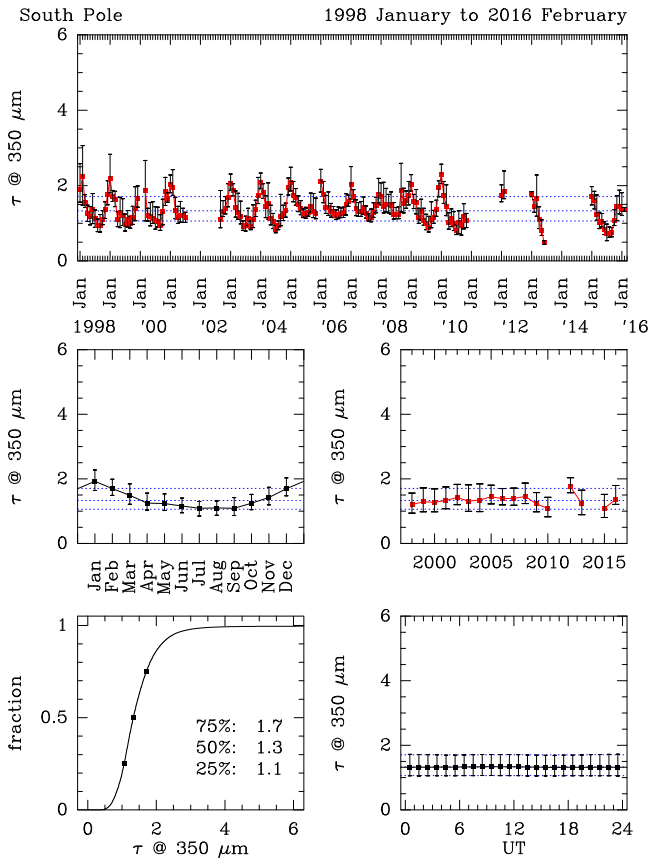}
\caption[sp-0-dat]{\label{fig:sp-0-dat}
Broad band 350\,$\mu$m zenith optical depth measured at the South Pole.
{\it Top:\/} Monthly quartiles (25\%, 50\%, \& 75\%); 
{\it center left:\/} seasonal variation;
{\it center right:\/} yearly quartiles;
{\it bottom left:\/} cumulative distribution;
{\it bottom right:\/} diurnal variation.
In each panel, markers indicate median values and the error bars show the 
first and third quartiles (25\% \& 75\%).
Horizontal dotted lines show the overall quartiles. 
}
\end{center}
\end{figure}

\begin{figure} 
\begin{center}
\includegraphics[width=0.99\columnwidth]{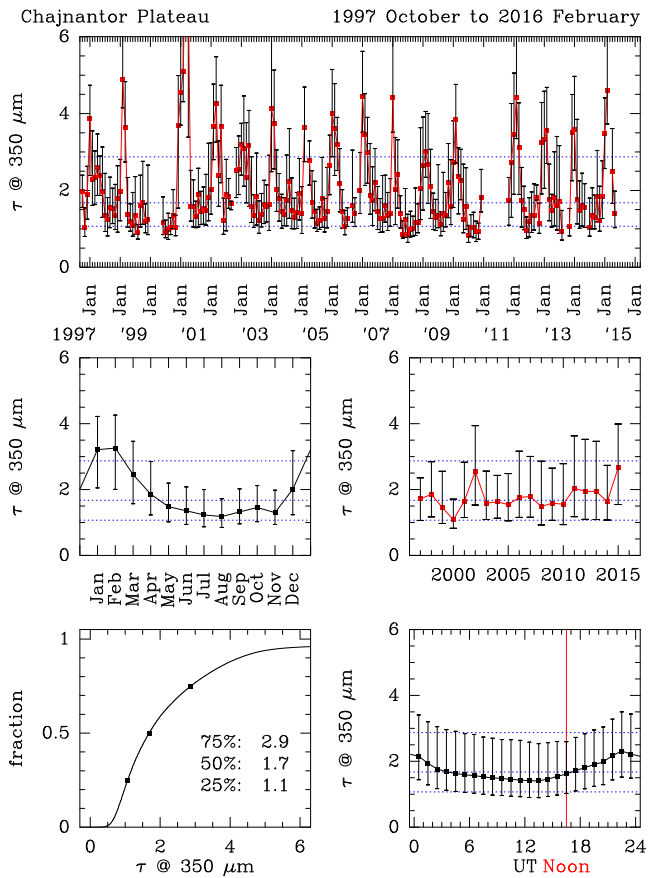}
\caption[cp-0-dat]{\label{fig:cp-0-dat}
Broad band 350\,$\mu$m zenith optical depth measured on the Chajnantor plateau. 
These data are a non redundant composite of measurements at ALMA, at the CBI, and 
at APEX.
{\it Top:\/} Monthly quartiles (25\%, 50\%, \& 75\%); 
{\it center left:\/} seasonal variation;
{\it center right:\/} yearly quartiles;
{\it bottom left:\/} cumulative distribution;
{\it bottom right:\/} diurnal variation (with mean solar noon indicated).
In each panel, markers indicate median values and the error bars show the 
first and third quartiles (25\% \& 75\%).
Horizontal dotted lines show the overall quartiles. 
} 
\end{center}
\end{figure}

\begin{figure} 
\begin{center}
\includegraphics[width=0.99\columnwidth]{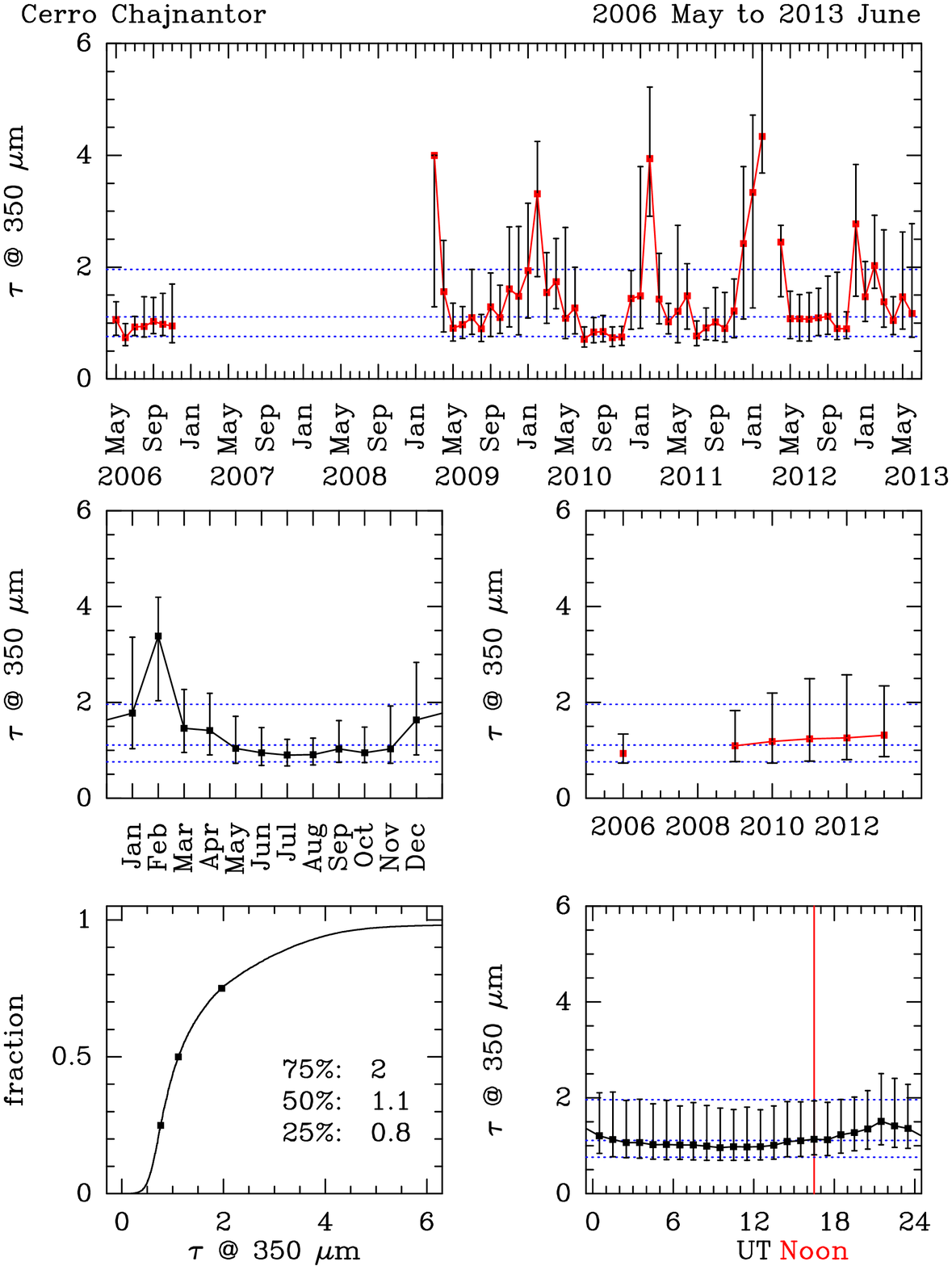}
\caption[cc-0-dat]{\label{fig:cc-0-dat}
Broad band 350\,$\mu$m zenith optical depth measured on Cerro Chajnantor.
{\it Top:\/} Monthly quartiles (25\%, 50\%, \& 75\%); 
{\it center left:\/} seasonal variation;
{\it center right:\/} yearly quartiles;
{\it bottom left:\/} cumulative distribution;
{\it bottom right:\/} diurnal variation (with mean solar noon indicated).
In each panel, markers indicate median values and the error bars show the 
first and third quartiles (25\% \& 75\%).
Horizontal dotted lines show the overall quartiles. 
} 
\end{center}
\end{figure}

\subsection{Variations}

At all locations, the atmospheric transparency is generally better 
during the winter and at night. 
On Maunakea (Fig.~\ref{fig:mk-0-dat}) 
the seasonal pattern is evident, if not prominent.
The seasonal contrast, calculated as the ratio of the 
maximum and minimum monthly median optical depths, is 1.5. 
Interannual variations are as strong as the seasonal pattern, 
both for the same month in different years and for entire years.
At the South Pole (Fig.~\ref{fig:sp-0-dat}), 
conditions are remarkably consistent from year to year.
The seasonal contrast is 1.8,
i.\,e., the optical depth during summer is almost twice
as large as during winter.
In the Chajnantor region (Figs.\ \ref{fig:cp-0-dat} \& \ref{fig:cc-0-dat}), 
conditions are consistently good
from April through December but 
deteriorate during the summer months
when a shift in atmospheric 
circulation draws moist air 
over the Andes from the Amazon basin.
The seasonal contrast is 2.6 on the Chajnantor plateau.
There is considerable interannual variation in the severity 
of the summer season; 
winter conditions are more consistent.
On Cerro Chajnantor, the seasonal contrast is 3.7, 
primarily because of better winter conditions. 

The transparency is better at night.
Both on Maunakea and in the Chajnantor area, the variation lags behind the sun;
the best conditions occur around sunrise
and the optical depth is highest in the afternoon.
This may reflect the influence of inversion layers, 
or at least humid layers, 
that rise during the day and then subside as the night progresses.
The diurnal contrast, calculated the same way as the seasonal contrast,
is 1.6 at Maunakea, 1.7 on the Chajnantor plateau, and 
1.5 on Cerro Chajnantor.
Diurnal variations in the Chajnantor area are less pronounced 
during the winter than during the summer.

None of the sites display any significant secular trend in the median optical depth
over the 18 year duration of the measurements.
At all three locations, 
correlations between 
the monthly median optical depths
and the monthly Multivariate ESNO Index (MEI; \citealp{wolter:2011})
are insignificant ($r < 0.1$).

\subsection{Cross Comparisons}

In Chile and on Maunakea, there were several opportunities for 
cross comparisons:
between two tippers at nearby locations, Cerro Chajnantor and the Chajnantor plateau;
between measurements made alternately at two wavelengths by a single tipper;
and with other colocated instruments.
In all cases, the cross comparison data are highly correlated. 

Outliers are more numerous in the tipper measurements 
than expected for a normal distribution.
To suppress the undue influence of these outliers,
the data ranges considered in the regressions 
were restricted.
In addition, data were rejected on the outskirts 
of the measurement distribution 
where the density of points 
fell below 2.5\% of the peak. 
This permits a robust characterization of the range
of most interest, good observing conditions.

\strut

\subsubsection{Cerro Chajnantor}

\begin{figure}
\begin{center}
\includegraphics[clip=true,
  width=0.45\columnwidth]
  {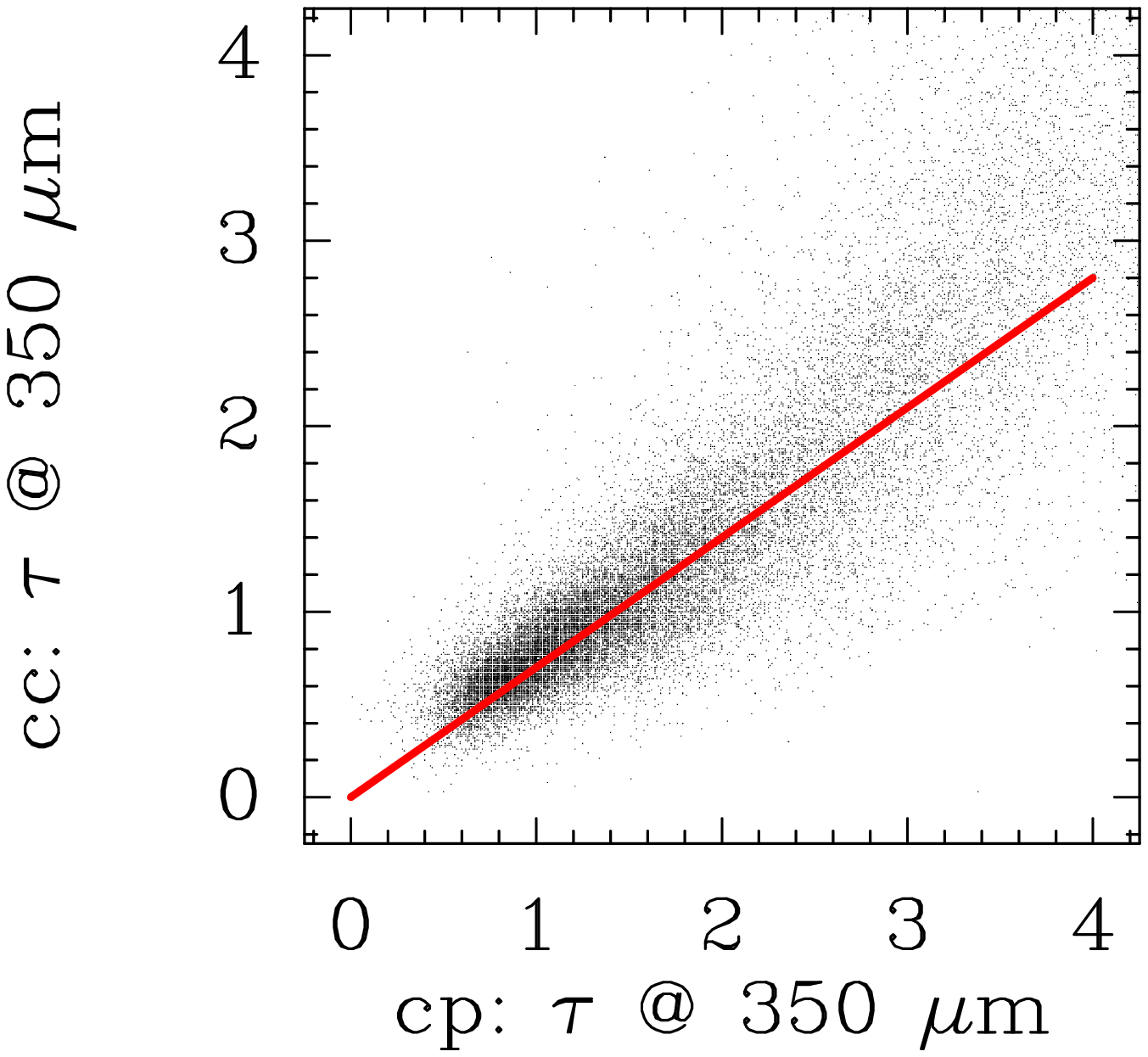}
\quad
\includegraphics[clip=true,
  width=0.49\columnwidth]
  {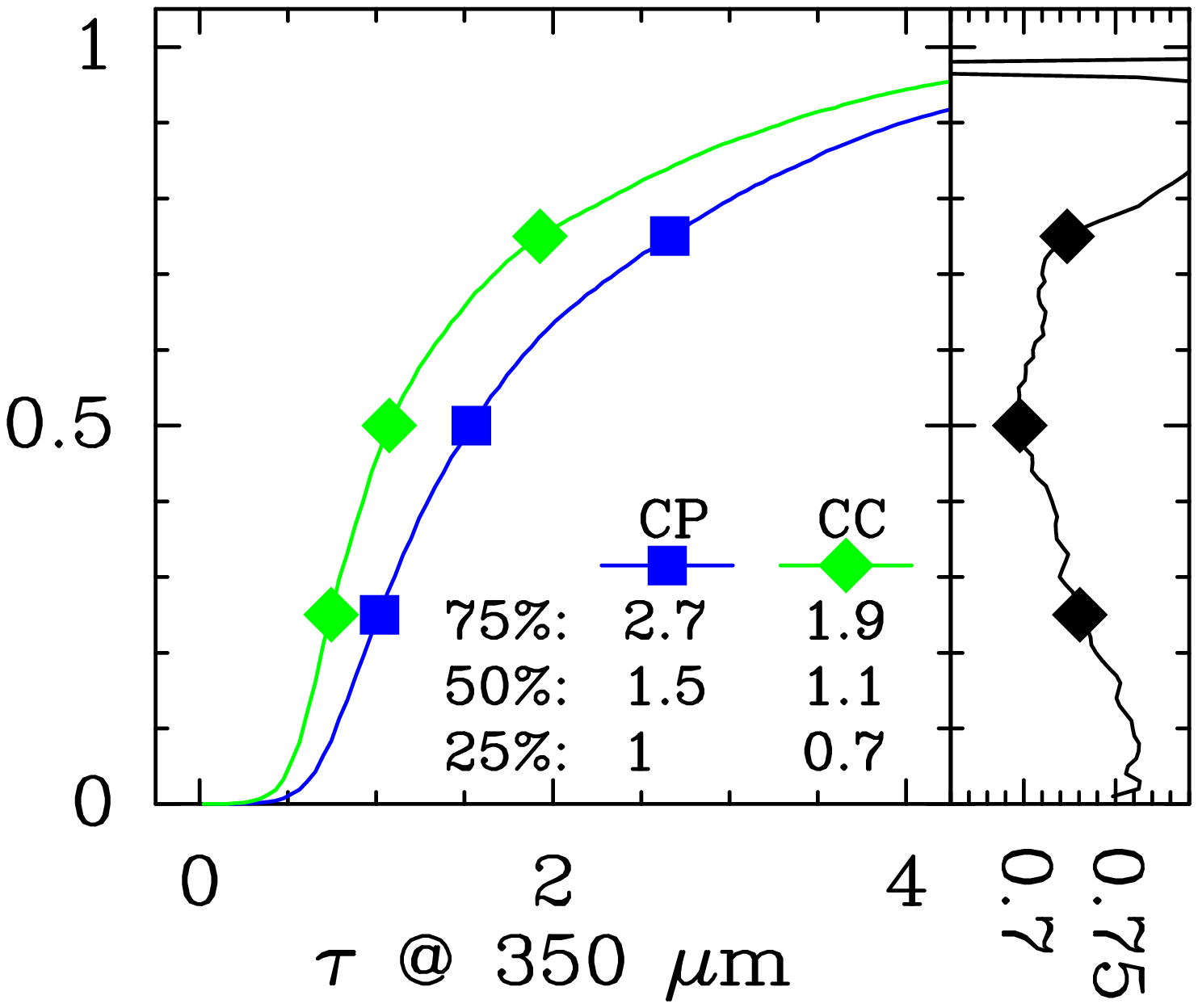}
\caption[cp-cc]{ \label{fig:cp-cc}   
{\it Left:\/}
Correlation between the broad band 350\,$\mu$m zenith optical depths measured 
simultaneously on the Chajnantor plateau (CP) and on Cerro Chajnantor (CC).
The guide line shows the best linear fit
$\tau_{\rm cc} = 0.7\, \tau_{\rm cp}$.
{\it Right:\/}
The cumulative distributions 
of the paired measurements and the ratio of the quantiles,
$Q_i (\tau_{\rm cc}) / Q_i (\tau_{\rm cp})$.
} 
\end{center}
\end{figure}
 
Because they are only 8\,km apart,
the two tippers
on Cerro Chajnantor and on the Chajnantor plateau
permit 
direct comparison of conditions with 
simultaneous measurements
(Fig.~\ref{fig:cp-cc}, {\it left\/}).
The paired measurements 
are highly correlated ($r > 0.9$) and 
indicate the 350\,$\mu$m atmospheric transparency 
is significantly better on Cerro Chajnantor 
than on the plateau.
The measurement distribution is curved, which
indicates the effects of saturation in the tipper measurements
under poor conditions.
For the best 65\% of conditions 
($\tau_{\rm cp} \le 2.05$ and $\tau_{\rm cc} \le 1.45$),
linear regression indicates
\begin{equation}
\tau_{\rm cc} = (0.7 \pm 0.05) \, \tau_{\rm cp} + (0.0 \pm 0.05) \ .
\end{equation}
Another indicator of relative conditions 
is the ratio of the quantiles for the two locations
(Fig.~\ref{fig:cp-cc}, {\it right\/}).
For the bulk (75\%) of the measurements,
this ratio is 0.7--0.75,
consistent with the regression parameters.
There is no significant seasonal or diurnal variation 
in the ratio.

As the altitude difference between the sites $\Delta h = 550$\,m, 
the 350\,$\mu$m optical depth ratio 
$ \tau_{\rm cc} / \tau_{\rm cp} = 0.7$ 
corresponds to an (exponential) scale height 
$ \Delta h / \ln (\tau_{\rm cp} / \tau_{\rm cc} ) = 1540$\,m.
As discussed below (\S 6.3), this optical depth scale height should not
be confused with the water vapor scale height.

\subsubsection{200\,$\mu$m transparency}

\begin{figure}
\begin{center}  
\includegraphics[clip=true, 
  width=0.45\columnwidth]
  {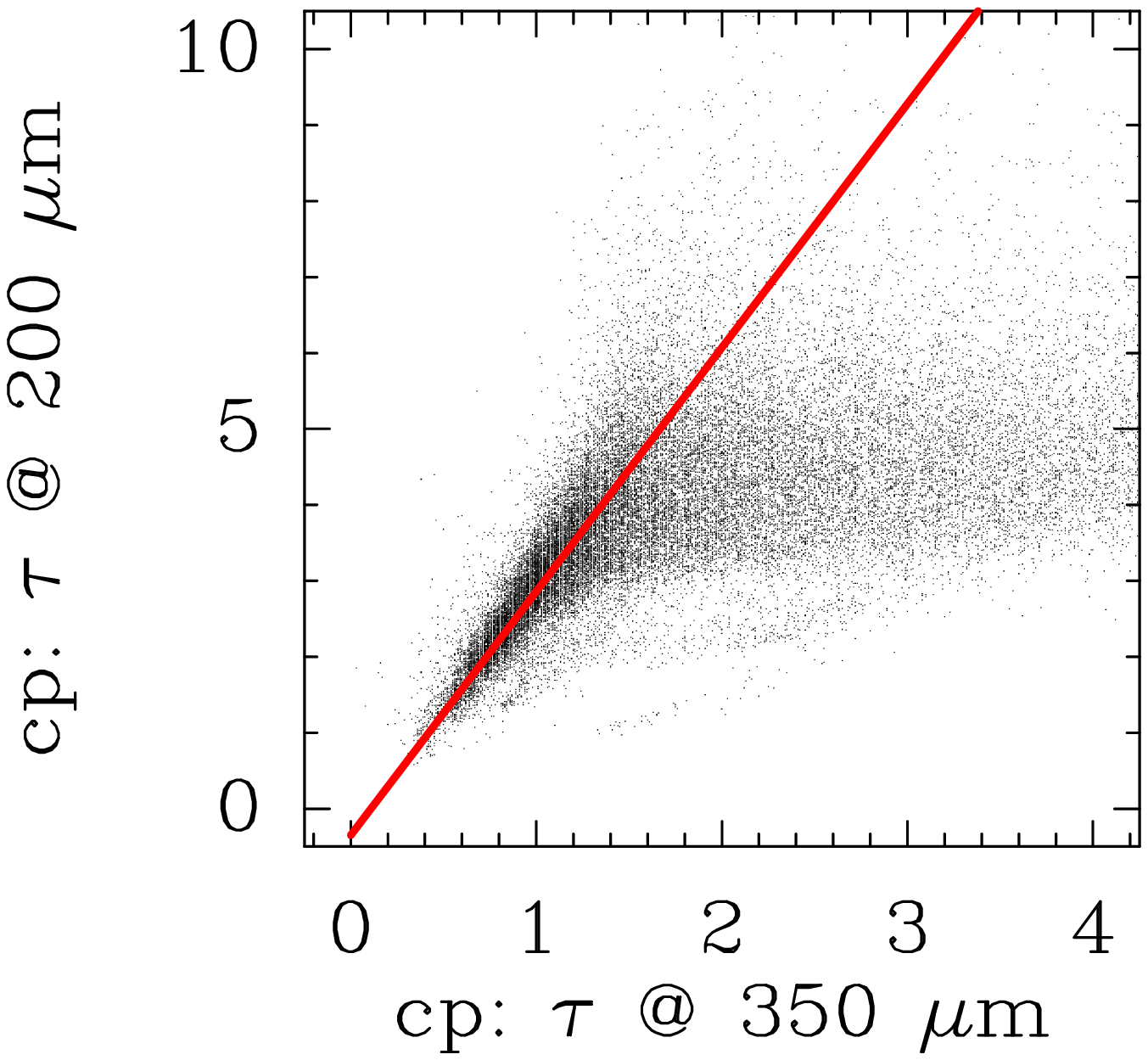}
\quad
\includegraphics[clip=true, 
  width=0.45\columnwidth]
  {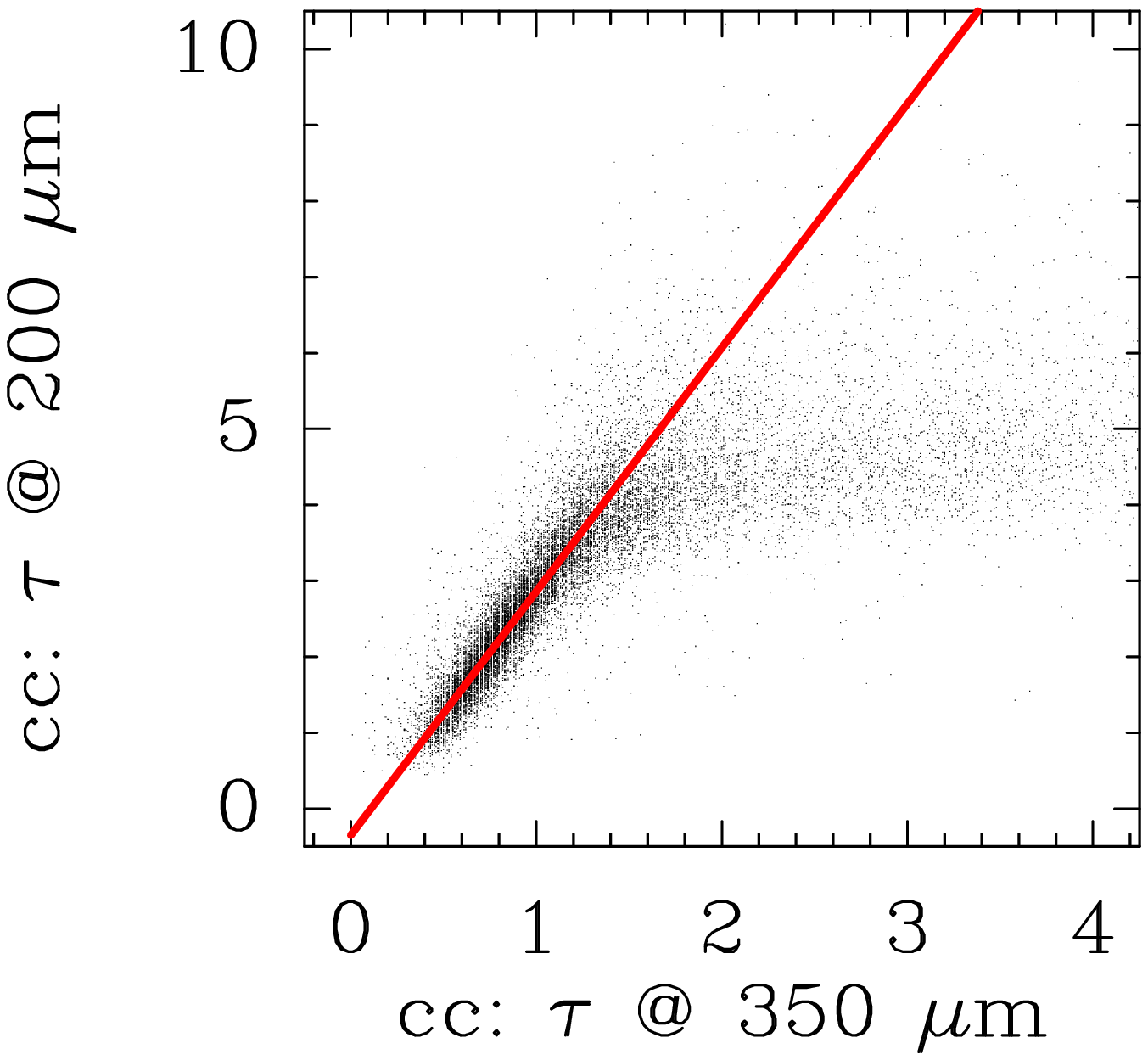} 
\caption[cmp-200]{ \label{fig:cmp-200}   
Correlations between successive measurements of the 
broad band 350\,$\mu$m and 200\,$\mu$m
zenith optical depths 
on the Chajnantor plateau 
({\it left\/})
and on Cerro Chajnantor
({\it right\/}).
The measurements saturate when $\tau(200\,\mu{\rm m}) > 4$.
The guide lines illustrate 
$\tau (200\,\mu{\rm m}) = 3.2 \,\tau (350\,\mu{\rm m}) - 0.35$.
} 
\end{center}
\end{figure}

\begin{figure}
\begin{center}
\includegraphics[clip=true,
  width=0.45\columnwidth]
  {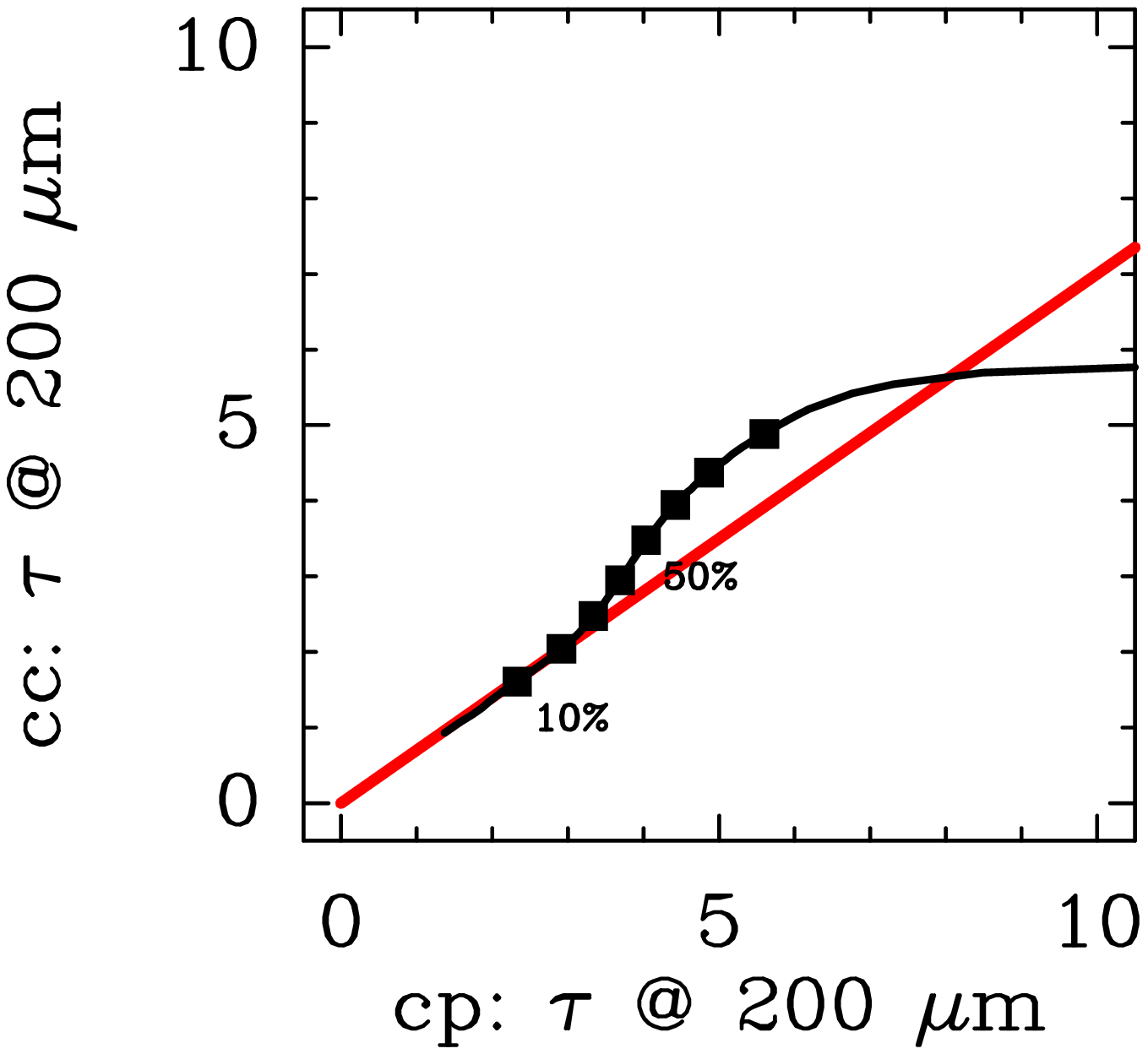} 
\quad
\includegraphics[clip=true,
  width=0.49\columnwidth]
  {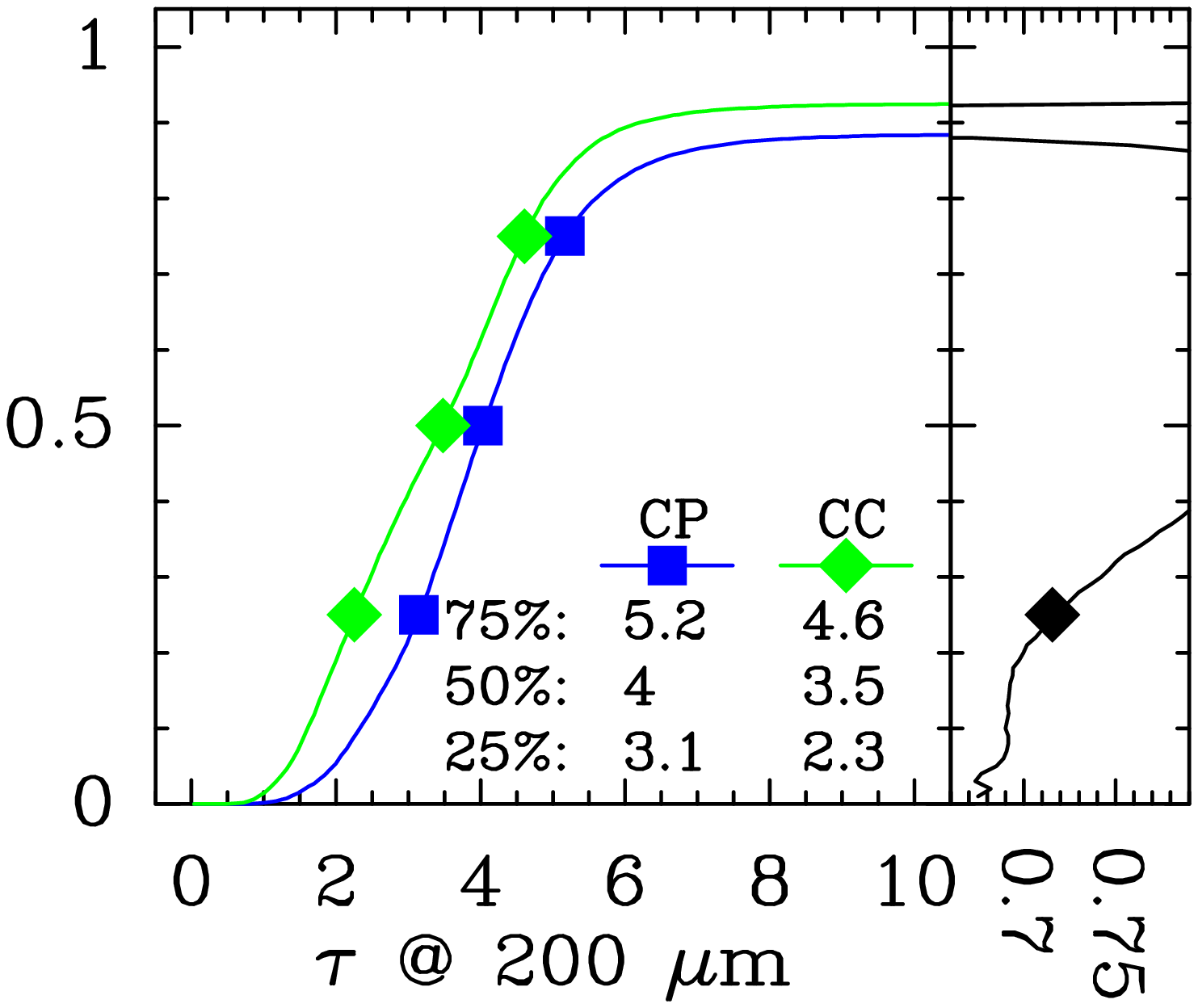} 
\caption[cp-cc-200]{ \label{fig:cp-cc-200}   
{\it Left:\/}
Quantiles of the 
broad band 200\,$\mu$m zenith optical depths
measured on the Chajnantor plateau (CP)
and on Cerro Chajnantor (CC) with the deciles marked 
($QQ$ plot).
The measurements were not simultaneous.
The guide line shows 
$\tau_{\rm cc} = 0.7\, \tau_{\rm cp}$.
{\it Right:\/}
The cumulative distributions 
of the 200\,$\mu$m  measurements and the ratio of the quantiles,
$Q_i (\tau_{\rm cc}) / Q_i (\tau_{\rm cp})$.
} 
\end{center}
\end{figure} 

Under exceptionally dry conditions, observations at 200\,$\mu$m may 
be contemplated
\citep{mankin:1973, matsushita:1999, paine:2000, blundell:2002, pardo:2005, ward:2005}.
One of the tippers in Chile was fitted with an additional bandpass filter 
so it alternates every 13\,min between a measurement at 350\,$\mu$m and 
one at 200\,$\mu$m.
Both on the Chajnantor plateau and on Cerro Chajnantor, 
there is a strong correlation between 
successive measurements at the two wavelengths
(Fig.~\ref{fig:cmp-200}).
Saturation of the 200\,$\mu$m measurements is clearly apparent under poor conditions,
when $\tau > 4$.
For good conditions, $\tau (350\,\mu{\rm m}) \le 1.55$, linear regression indicates
\begin{equation}
\tau (200\,\mu{\rm m}) = (3.2 \pm 0.3) \,\tau (350\,\mu{\rm m}) - (0.35 \pm 0.3) \ ,
\end{equation}
with no significant difference between the two sites. 
This measured optical depth ratio is similar to previous determinations 
\citep{matsushita:1999, ward:2005}.

In the absence of simultaneous 200\,$\mu$m measurements at the two locations,
comparing the distributions of the measurements
provides an indication of relative site quality
despite the risk of selection bias.
Under good conditions 
when the measurements are not badly saturated, 
$\tau (200\,\mu{\rm m})< 3.5$ (first quartile and better),
the optical depth ratio is 70\% (Fig.~\ref{fig:cp-cc-200}), 
the same as for the 350\,$\mu$m measurements.

Even at excellent sites, 200\,$\mu$m  observations 
will be only practical under exceptional circumstances
because the atmospheric transmission is otherwise so poor. 
On Cerro Chajnantor, for example, the first quartile zenith optical 
depth is 2.3, which corresponds to 10\% transmission.
Exceptional observing conditions occur
more often on Cerro Chajnantor than on the plateau.

\subsubsection{183\,GHz water vapor line}

\begin{figure}
\begin{center}
\includegraphics[clip=true,
  width=0.45\columnwidth]
  {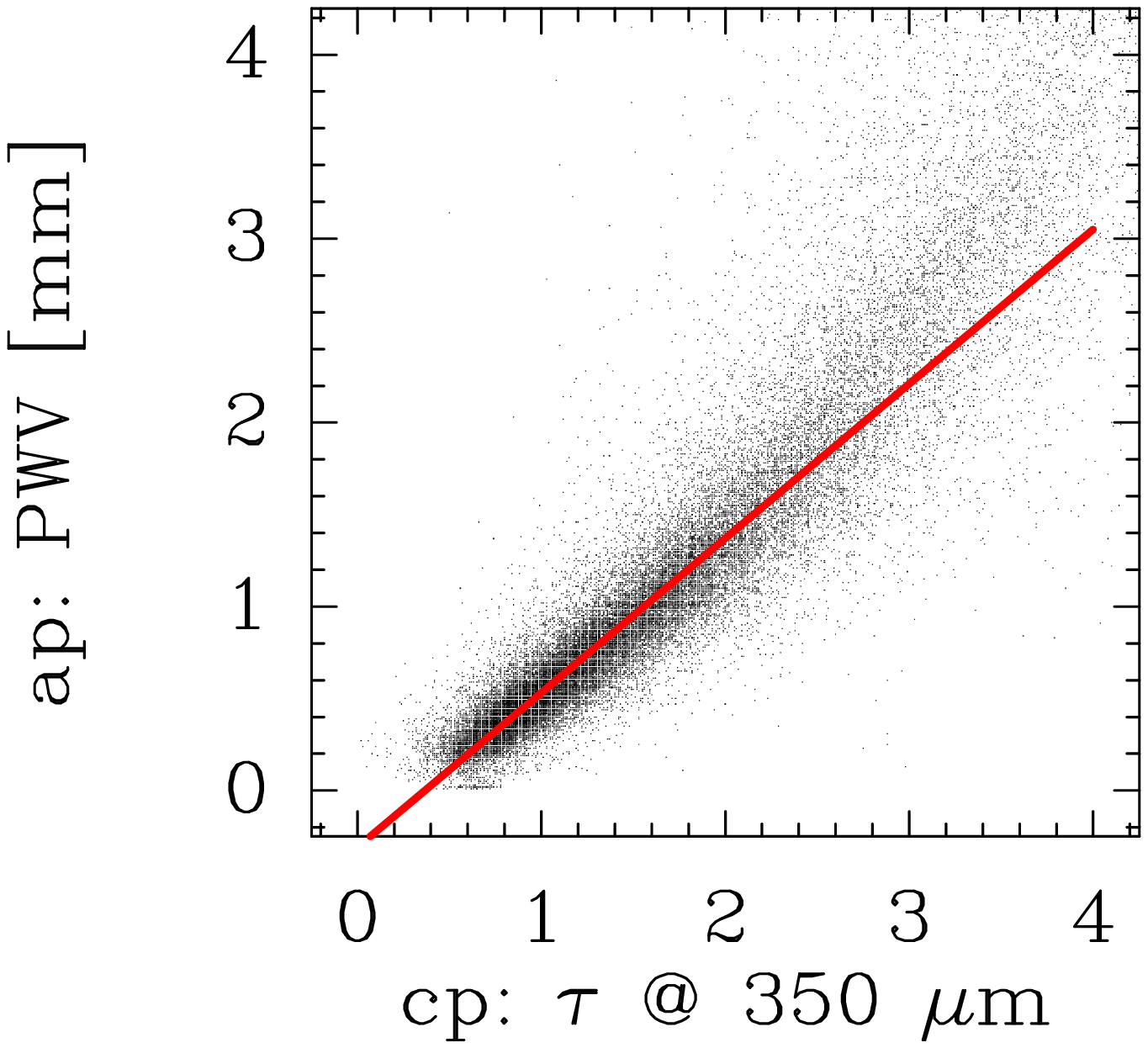} 
\quad
\includegraphics[clip=true,
  width=0.45\columnwidth]
  {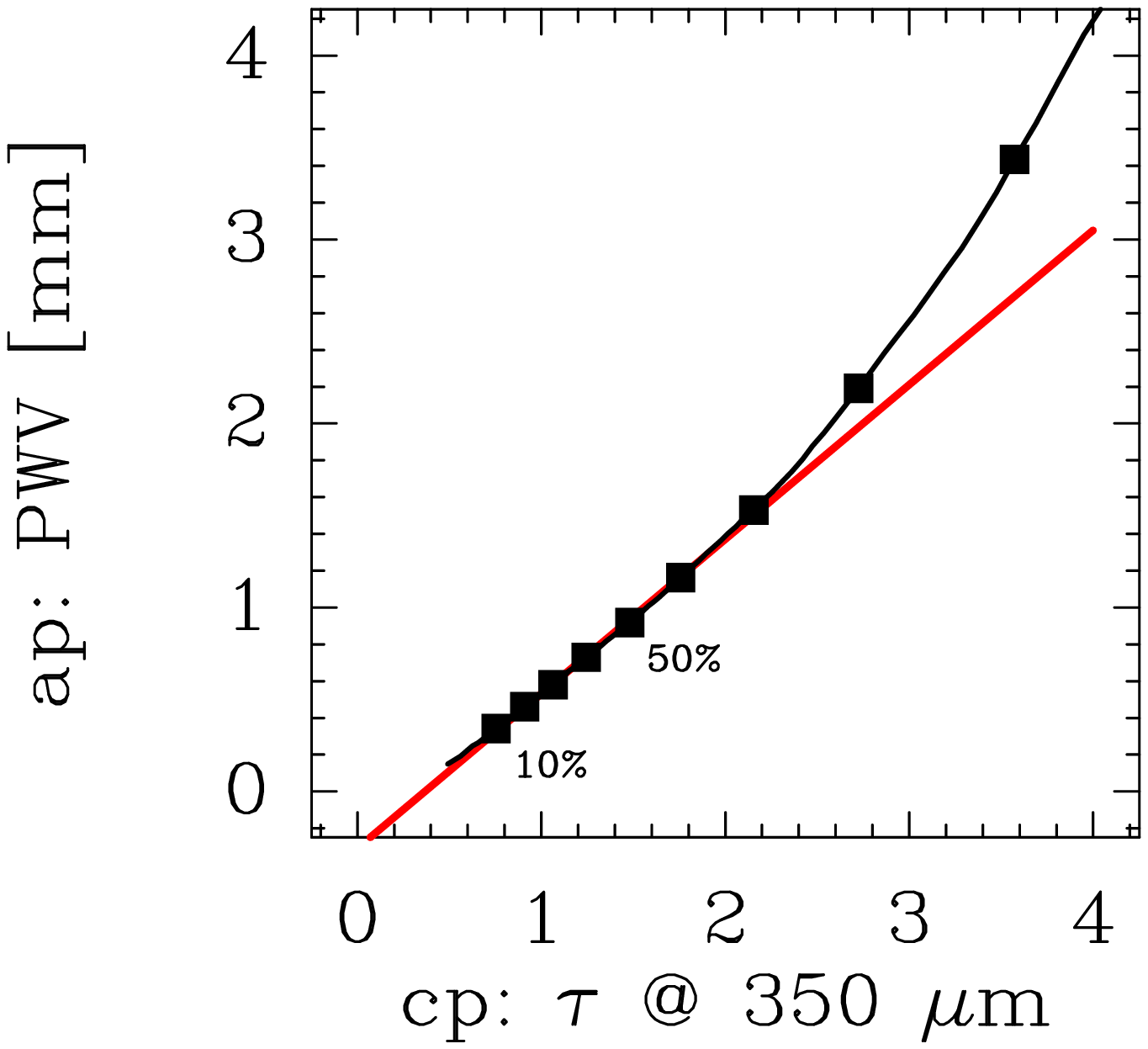} 
\caption[cp-pwv]{ \label{fig:cp-pwv}   
{\it Left:\/}
Correlation between simultaneous measurements of the 
broad band 350\,$\mu$m zenith optical depth on 
the Chajnantor plateau 
and the precipitable water vapor (PWV) column density
derived from simultaneous 183\,GHz spectroscopic measurements
at APEX.
{\it Right:\/}
Quantiles of the paired measurements 
with the deciles marked
($QQ$ plot).
In both panels, 
the guide line illustrates 
PWV [mm]${} = 0.84\, \tau(350\,\mu{\rm m}) -0.31$.
} 
\end{center}
\end{figure} 

Radiometry of spectral lines, including the 183\,GHz line, is a standard technique for 
measuring the atmospheric water vapor content (see \citealp{askne:1986}).
Furthermore, the correspondence between water vapor and submillimeter atmospheric 
transparency is well established for individual sites
(e.\,g., \citealp{matsushita:2003, pardo:2004, tamura:2011}).
On the Chajnantor plateau, a heterodyne spectrometer was mounted 
on the APEX telescope in 2006 to measure the strength of 
the 183\,GHz line
during astronomy observations.
These independent measurements of the 
precipitable water vapor (PWV)
column density 
are well correlated with
simultaneous tipper measurements
of the 350\,$\mu$m optical depth
(Fig.~\ref{fig:cp-pwv}, {\it left\/}).
Under most (70\%) conditions, $\tau(350\,\mu{\rm m}) < 2.2$, 
the correlation is linear
but there is a significant departure from linearity under poor (30\%) conditions,
indicating saturation in the 350\,$\mu$m measurement.
The quantiles of the paired measurements provide another indication of
the correlation (Fig.~\ref{fig:cp-pwv}, {\it right\/}).
Tracing the ridge along the maximum of the measurement distribution,
the quantiles clearly show the linear correlation under good conditions 
and the curvature of the distribution under poor conditions.
For $\tau(350\,\mu{\rm m}) < 2.55$, linear regression indicates
\begin{equation}
{\rm PWV\ [mm]} = 0.84\, \tau(350\,\mu{\rm m})  - 0.31 \ .
\end{equation} 
Although 
this particular relation between PWV and $\tau(350\,\mu{\rm m}$)
is only valid in the environs of Chajnantor (\S 6.2),
the measured correlation reaffirms that the tipper data are good 
indicators of observing conditions at all sites.

Extrapolating the measurements to the limit of no water vapor, 
the optical depth would be $\tau(350\,\mu{\rm m}) = 0.38\ (= 0.31/0.84)$.
Although artifacts in the tipper measurements or in the radiometer measurements
(or in both) may contribute,
this zero PWV optical depth is indicative of the  
absorption caused by dry air
(see \citealp{pardo:2001a, pardo:2001b, pardo:2005}).

\subsubsection{225\,GHz transparency}

\begin{figure}
\begin{center}
\includegraphics[clip=true,
  width=0.45\columnwidth]
  {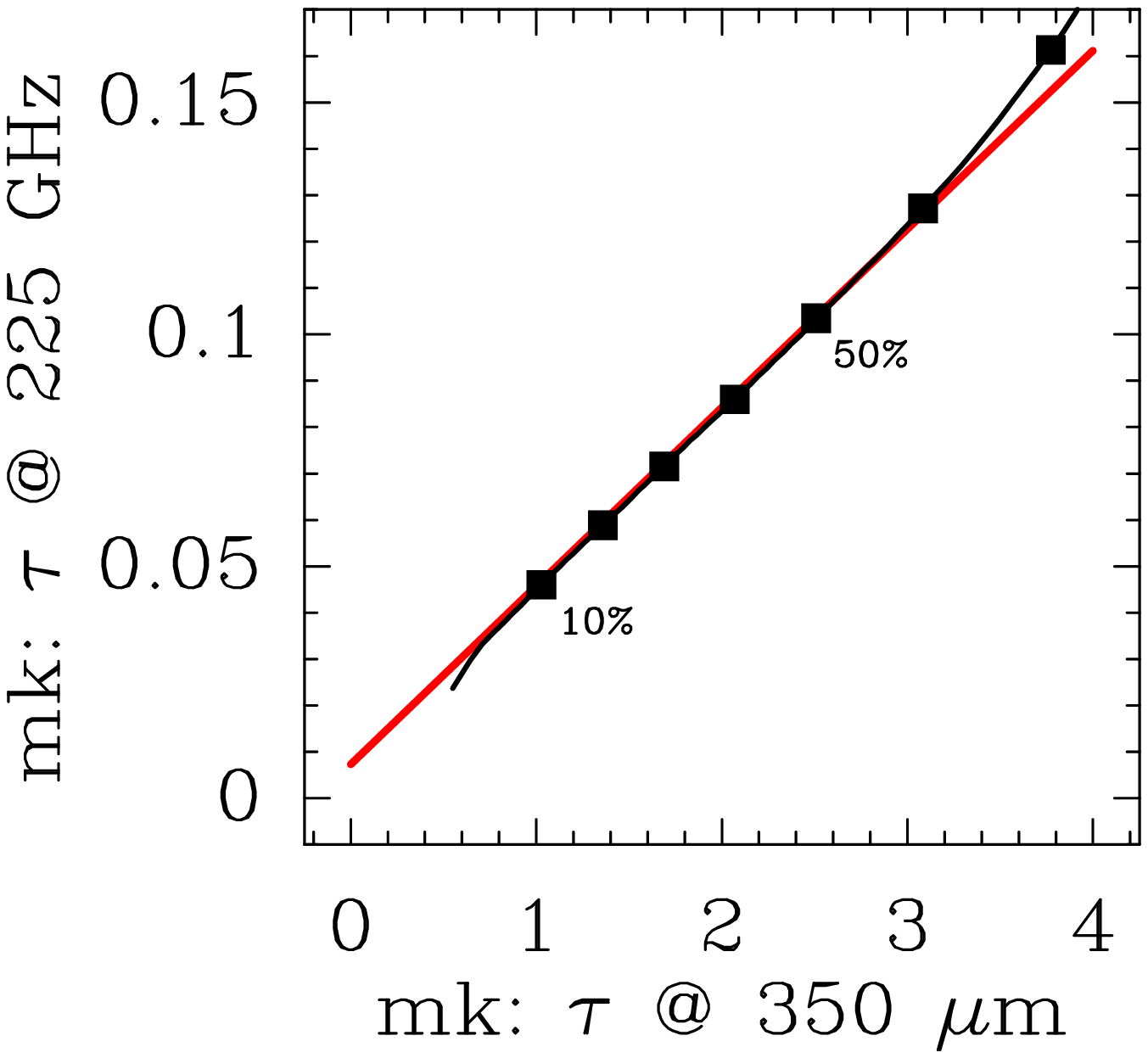}
\quad
\includegraphics[clip=true,
  width=0.45\columnwidth]
  {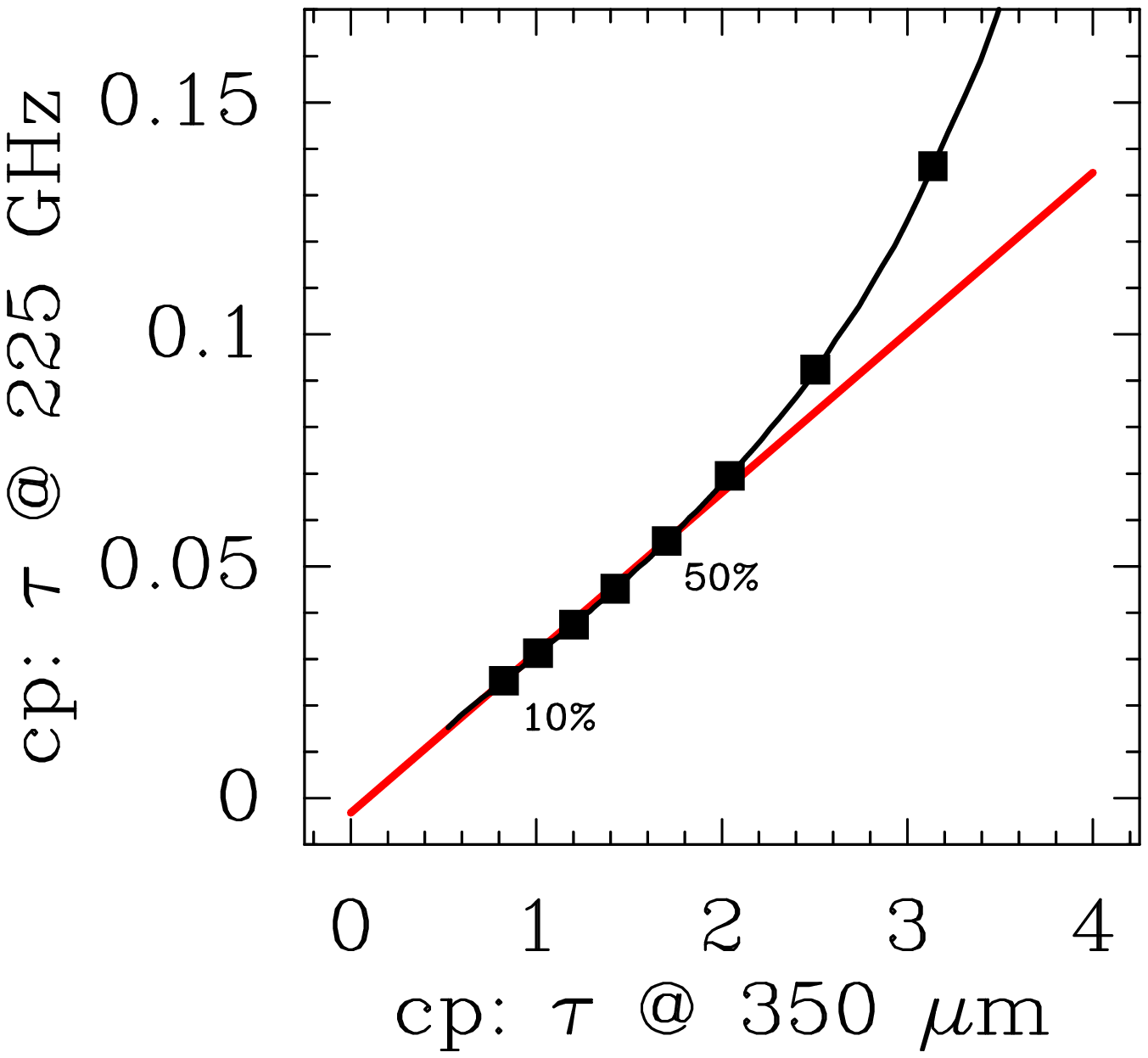}
\\
\strut
\\
\includegraphics[clip=true,
  width=0.45\columnwidth]
  {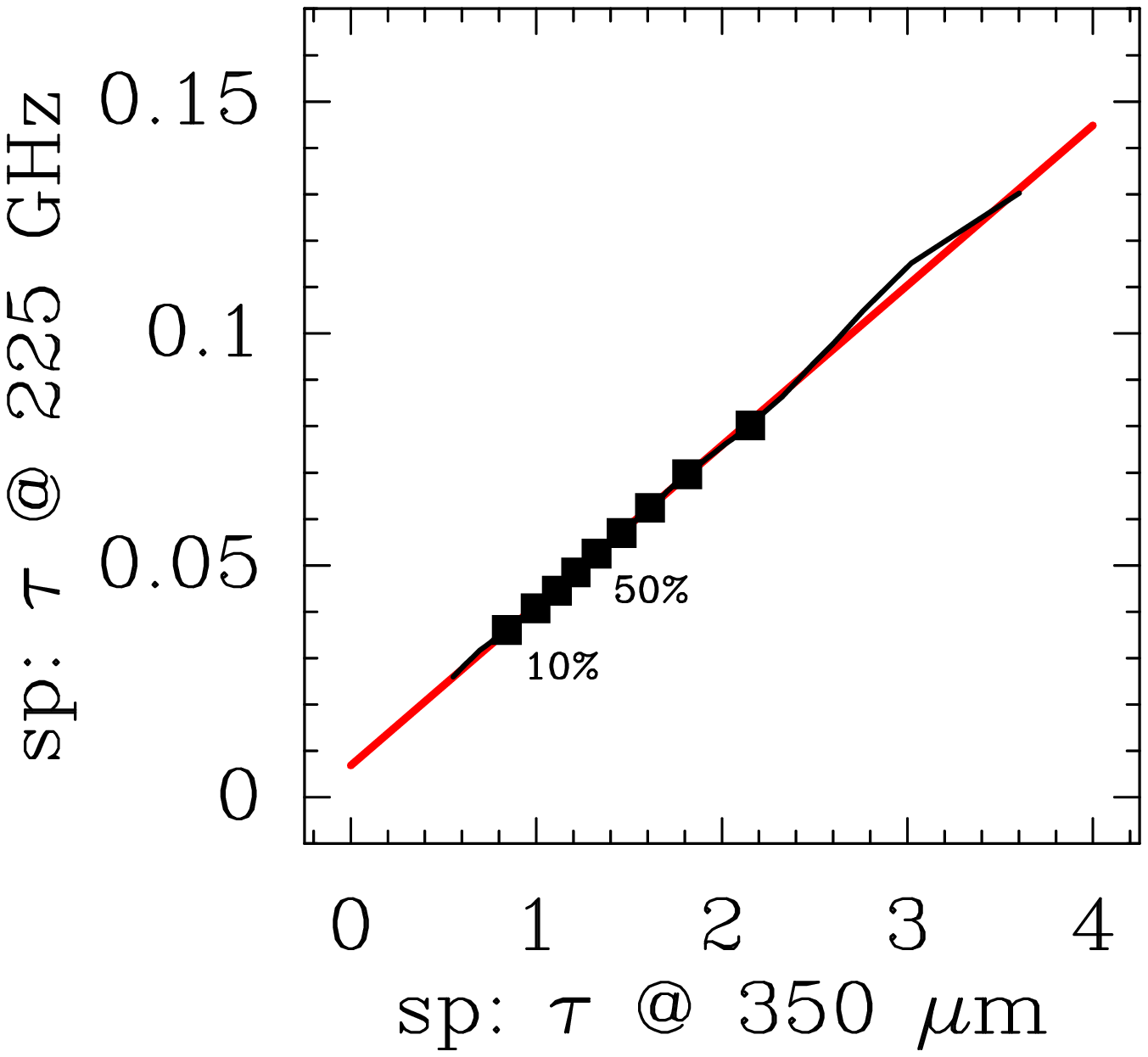}
\caption[cmp-225]{ \label{fig:cmp-225}   
Quantiles of  
the broad band 350\,$\mu$m
and the narrow band 225\,GHz
zenith optical depths measured
simultaneously on Maunakea ({\it top left\/}), 
simultaneously on the Chajnantor plateau ({\it top right\/}),
but many years apart at the South Pole ({\it below\/})
($QQ$ plots). Deciles are marked.
The guide lines illustrate
$\tau(350\,\mu{\rm m}) = 26\, \tau(225\,{\rm GHz}) -0.2 $ for Maunakea,
$\tau(350\,\mu{\rm m}) = 29\, \tau(225\,{\rm GHz}) +0.1 $ for Chajnantor, and
$\tau(350\,\mu{\rm m}) = 29\, \tau(225\,{\rm GHz}) -0.2 $ for the South Pole.
} 
\end{center}
\end{figure} 

Transparency measurements at or about 225\,GHz have been
made to characterize observing conditions at many sites
(e.\,g., \citealp{hogg:1988, kohno:1995}).
On Maunakea and on the Chajnantor plateau,
preexisting narrow band 225\,GHz heterodyne tippers 
\citep{radford:2000}
operated simultaneously with the submillimeter tippers.
At both locations, the measured broad band 350\,$\mu$m 
and narrow band 225\,GHz zenith optical depths are well correlated
(Fig.~\ref{fig:cmp-225}).
On Maunakea when $\tau(350\,\mu{\rm m})  < 2.55$, linear regression indicates
\begin{equation}
\tau(350\,\mu{\rm m}) = (26 \pm 3)\, \tau({\rm 225\,GHz}) - (0.2 \pm 0.2) 
  \phantom{\ ,}
\end{equation}
and on the Chajnantor plateau when $\tau(350\,\mu{\rm m})  < 2.05$,
\begin{equation}
\tau(350\,\mu{\rm m}) = (29 \pm 3)\, \tau({\rm 225\,GHz}) + (0.1 \pm 0.1) \ .
\end{equation}
The relatively large parameter uncertainties are caused by the scatter in both sets of measurements.

At the South Pole, a 225\,GHz tipper was operated in 1992
\citep{chamberlin:1994}, 
prior to the development of the submillimeter tipper. 
Although the measurements were made many years apart, because 
the interannual consistency at the South Pole comparison of the 
distribution quantiles provides an indication of the correspondence,
\begin{equation}
\tau(350\,\mu{\rm m}) = 29 \, \tau({\rm 225\,GHz}) - 0.2 \ .
\end{equation}

At least within the parameter uncertainties, 
the regression slopes for the three sites are consistent
and similar to previous, 
independent determinations near Chajnantor
\citep{matsuo:1998,matsushita:1999}.
The regression offsets, however, show a small but noticeable difference, 
corresponding to $\Delta\, \tau({\rm 225\,GHz}) \approx 0.01$.
This might be ascribed, at least partly, to systematic differences in 
the 225\,GHz tippers. 
Although they share a common original design \citep{hogg:1988}, 
they were modified and maintained independently 
over the years.
During the period of the submillimeter measurements, 
the 225\,GHz tippers were never compared side by side.
An incident on Maunakea suggests these instruments may be sensitive to 
setup conditions.
In 2015 January, the alignment of the scanning mirror on the 
225\,GHz tipper was damaged during an
ice storm. After repairs and realignment, the correlation between
subsequent 225\,GHz and 350\,$mu$m measurements exhibited a similar slope
but the offset decreased in magnitude
\citep{radford:2016}.
This incident highlights the limitations of these particular 
225\,GHz tippers for measurements under very dry conditions.

\section {Discussion}

Recent years have seen the development of good models of
radiative transfer in the atmosphere 
at microwave and submillimeter wavelengths, 
including, but not limited to, 
ATM \citep{pardo:2001a} and {\it am\/} \citep{paine:2014}.
(For the purposes of this discussion, the predictions of ATM and {\it am\/}
are essentially the same.)
These models provide a framework for interpreting the 
tipper measurements 
but do not, of course, 
predict how often good conditions might occur at any site.

\subsection{Bandwidth}

\begin{figure} 
\begin{center}
\includegraphics[clip=true, width=0.45\columnwidth] {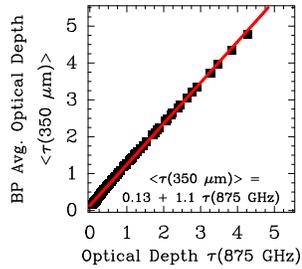}
\caption[narrow-wide]{\label{fig:narrow-wide}
Predicted linear correspondence between the broad band 350\,$\mu$m
optical depth, $\langle \tau(350\,\mu{\rm m})\rangle$,
and the narrow band 875\,GHz optical depth, $\tau(875\,{\rm GHz})$.
Markers show the predictions of
ATM \citep{pardo:2001a} for
the wide range of conditions 
encountered at the four deployment sites,
223\,K${} \le T_{\rm base} \le 293$\,K, 0\,mm${} \le {\rm PWV} \le 2$\,mm,
and 2835\,m${} \le {\rm altitude} \le 5000$\,m. 
The model calculations used standard atmospheric profiles with 
$T_{\rm base}$ as the ground level (base) air temperature.
The solid line and equation show a linear fit to the model predictions.
The particular regression coefficients depend on the reference frequency,
875\,GHz in this case.
} 
\end{center}
\end{figure}

Because the tippers incorporate broad band filters,
the measured zenith optical depth corresponds to an average of 
the atmospheric transmission spectrum, $\theta(\nu)$ 
weighted by the filter bandpass, $F(\nu)$,
\begin{equation}
\langle \tau\rangle = - \ln  \textstyle 
\left[  \int \theta(\nu)\, F(\nu)\, d\nu \middle/ \int F(\nu)\, d\nu \right] \ .
\end{equation} 
The average broad band transparency is less than the peak transparency 
in the center of the atmospheric window,
raising the question whether the broad band optical depth is 
a good diagnostic of conditions for narrow band observations.
Model calculations with ATM \citep{pardo:2001a} 
for a wide range of conditions
predict, however, an almost linear relationship
between the average broad band optical depth and 
the narrow band optical depth at a particular frequency
(Fig.~\ref{fig:narrow-wide}). 
Hence the measured broad band optical depth is 
a good diagnostic of conditions for both
broad band and narrow band observations.

\subsection{Atmospheric Transmission and Water Vapor}

\begin{figure} 
\begin{center}
\includegraphics[width=0.7\columnwidth]{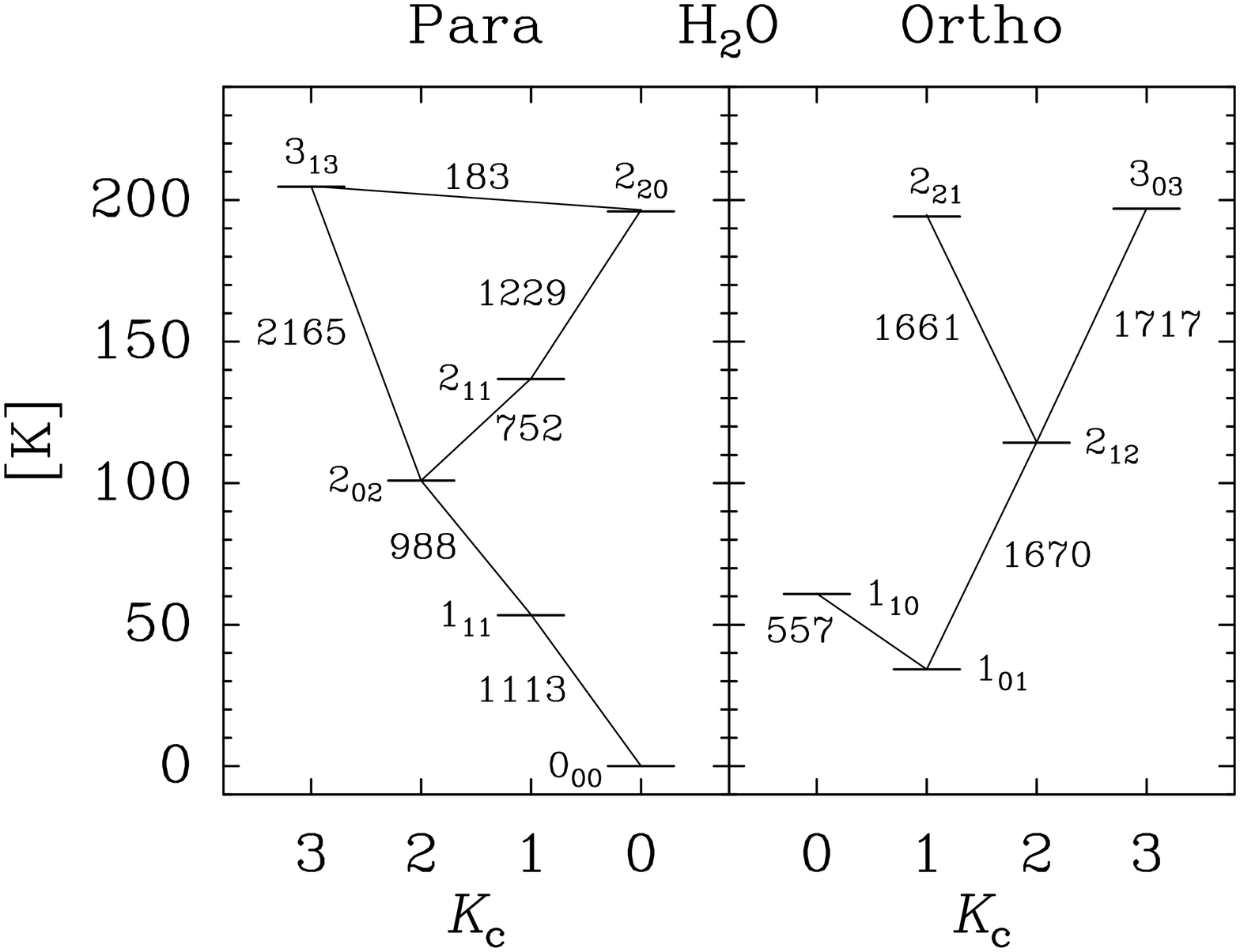}
\caption[water]{\label{fig:water}
Energy levels of low excitation water vapor rotational states with 
selected dipole transition frequencies denoted in GHz (after \citealt{chantry:1971}).
} 
\end{center}
\end{figure}

\begin{figure} 
\begin{center}
\includegraphics[width=0.99\columnwidth]{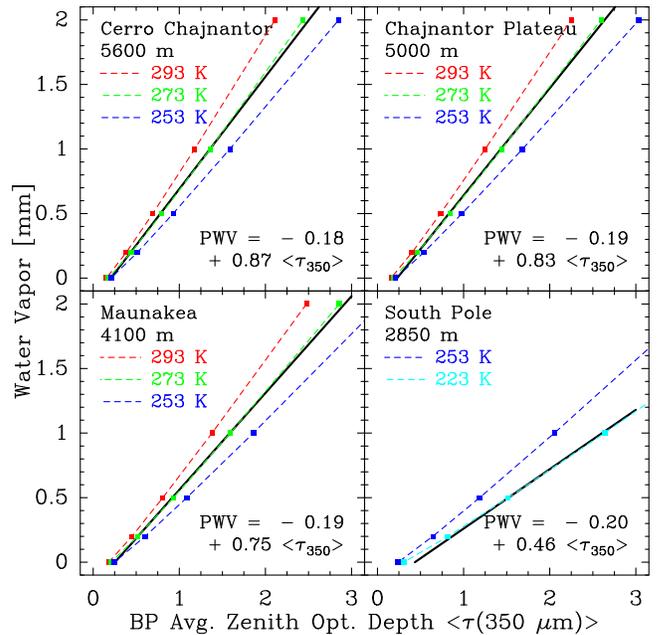}
\caption[pwv-350u]{\label{fig:pwv-350u}
Predicted correspondence between 
broad band 350\,$\mu$m optical depth and
precipitable water vapor (PWV)
for the range of temperatures and altitudes
characteristic of the four deployment sites.
The markers and dashed lines show the predictions of
ATM \citep{pardo:2001a} while 
the solid lines and equations indicate linear 
fits to those model predictions for 273\,K on Maunakea and Chajnantor
and for 223\,K at the South Pole.
}
\end{center}
\end{figure}

\begin{figure} 
\begin{center}
\includegraphics[width=0.99\columnwidth]{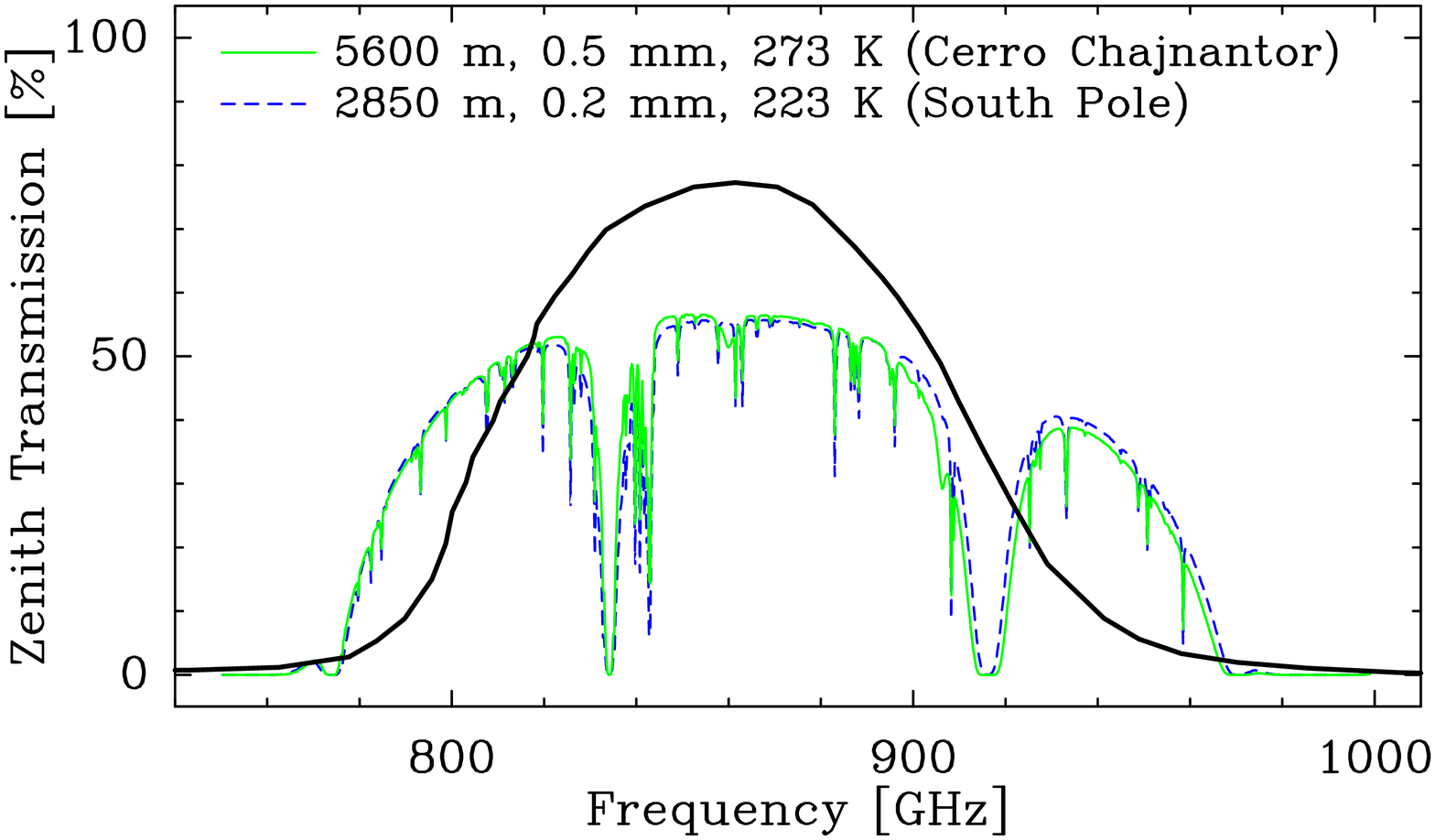}
\caption[model-compare]{\label{fig:model-compare}
Atmospheric transmission across the 350\,$\mu$m window
predicted by ATM \citep{pardo:2001a}
for good conditions at the South Pole (0.2\,mm PWV) and
on Cerro Chajnantor (0.5\,mm PWV).
The measured transmission spectrum of the tipper's 
bandpass filter is overlaid ({\it solid line\/}).
}
\end{center}
\end{figure}

Because water vapor has such a large effect on atmospheric transmission,
the precipitable water vapor (PWV)
is often used as a metric of observing conditions,
especially when considering time variations at a single site.
When comparing sites with very different circumstances, however,
it can be deceptive to focus solely on the PWV.

The submillimeter atmospheric windows are bracketed by strong, 
pressure broadened, and saturated transitions between 
low excitation water vapor rotational states
(Fig.~\ref{fig:water}). 
Transmission in these windows depends primarily on three quantities:
the pressure, determined by the site altitude; the air temperature; 
and the the water vapor column density (PWV).
Models of atmospheric radiative transfer
predict the windows are more transparent 
at higher altitude,
at higher temperature, 
and with less water vapor
(Fig.~\ref{fig:pwv-350u}).
The effects of water vapor and altitude (pressure) have straightforward explanations. 
When there is more water vapor, the lines are stronger and more saturated  
so the transmission is lower. 
At higher altitude (lower pressure), there is less pressure broadening so the line wings 
intrude less on the windows and the transmission is higher. 

The temperature effect is perhaps less intuitive. 
On the one hand, 
saturated cold air holds less moisture than saturated warm air. 
In Antarctica, the absolute humidity is very low
even though the very cold air in close contact with the ice sheet
is almost saturated, with a typical relative humidity around 80\%
\citep{gettelman:2006}.
At Chajnantor or Maunakea, by contrast, the best conditions occur
occur when the air is unsaturated and the relative humidity is very low, 
around 10\% or less.
On the other hand, the major water vapor absorption 
lines link states with excitation temperatures of 50--150\,K 
(Fig.~\ref{fig:water}). 
For the same amount of water vapor, 
the populations of these low excitation states will be greater 
at lower temperature than at higher temperature. 
This has the same effect as increasing the humidity: 
the lines are stronger and more saturated so the transmission is lower.
The temperature--humidity tradeoff can be dramatic. 
For good winter conditions at the South Pole and 
on Cerro Chajnantor, 
a 50\,K temperature difference
is roughly equivalent to doubling the amount of water vapor 
(Fig.~\ref{fig:model-compare}).

For the Chajnantor plateau, the slopes of the 
predicted correspondence between optical depth and PVW
(Fig.~\ref{fig:pwv-350u})
and of the measured correlation between 
the tipper and 183\,GHz radiometer measurements
(Fig.~\ref{fig:cp-pwv} \& Eq.~7)
are very close.
But the predicted offset, $-0.19$\,mm, is only two thirds of 
the measured offset, suggesting a
residual instrumental artifact
in the tipper or the radiometer or both.

\subsection{Water Vapor Scale Height}

\begin{deluxetable}{lccccc}
\tablewidth{0pt}
\tablecaption{\label{tab:wv-height}
\strut Water Vapor Scale Height 
} 
\tablehead{
  & \multicolumn{2}{c}{$\tau(350\,\mu{\rm m})$}& \multicolumn{2}{c}{PWV [mm]}& WV\\
  & Chaj.& Cerro& Chaj.& Cerro& scl.\,ht.\\
  & plateau& Chaj.& plateau& Chaj.& [m]\,\tablenotemark{*}}
  \strut
\startdata
75\,\%& 2.7& 1.9& 2.0\phantom{0}& 1.3\phantom{0}& 1280\\
50\,\%& 1.5& 1.1& 1.0\phantom{0}& 0.6\phantom{0}& 1080\\
25\,\%& 1.0& 0.7& 0.53& 0.28& \phantom{0}860
\enddata
\tablenotetext{*}{WV scale height $= 550\,{\rm m} / \ln ({\rm PWV}_{\rm cp} / 
  {\rm PWV}_{\rm cc} )$ }
\end{deluxetable}

Cerro Chajnantor rises 550\,m above the Chajnantor plateau. 
The measured optical depth ratio between these locations
(Fig.~\ref{fig:cp-cc})
corresponds to a scale height of 1540\,m (\S 5.3.1).
Because
of the zero water vapor offset in the optical depth measurements, 
however,
which is at least partly due to dry air absorption, 
the optical depth and water vapor scale heights are not the same.
Moreover, the optical depth ratio is constant, 
$\tau_{\rm cc} / \tau_{\rm cp} = 0.7$,
for the majority of the measurements.
Hence the water vapor scale height is {\it not\/} constant, but
varies with conditions: it is smaller when conditions are good. 

To estimate the typical water vapor scale height,
the measured optical depth ratio 
(Fig.~\ref{fig:cp-cc} \& Eq.~5) may be combined with
the measured correlation between the optical depth and water vapor 
on the Chajnantor plateau.
(Fig.~\ref{fig:cp-pwv} \& Eq.~7).
The resulting median ratio, 
${\rm PVW}_{\rm cc} / {\rm PVW}_{\rm cp} = 0.6$,
corresponds to an
(exponential) water vapor scale height 
of 1080\,m (Table~\ref{tab:wv-height}).
The first quartile scale height is smaller, 860\,m, 
and the third quartile is larger, 1280\,m.
This calculation ignores, however, the predicted altitude dependence
of the correspondence between optical depth and
water vapor 
(Fig.~\ref{fig:pwv-350u}).
If the predicted correspondence is
used, rather than the measured correlation,
then 
the estimated water vapor scale height is about 20\% larger,
with a median about 1300\,m.

These estimates of the water vapor scale height
and PWV ratio
are similar to previous results 
for the Chajnantor vicinity.
Radiosondes launched in 1998--2000 from the Chajnantor plateau 
indicated a typical water vapor scale height of 1.1\,km
\citep{giovanelli:2001}.
Sequential submm FTS measurements on the Chajnantor plateau in 1997--1999 and 
40\,km north on Cerro Sairecabur (5500\,m) in 2000--2002
found a typical PWV ratio of 0.6
\citep{blundell:2002}.
Simultaneous 183\,GHz water vapor line measurements 
on Cerro Chajnantor and at APEX in 2011 December  
indicated a typical PWV ratio of 0.64
\citep{bustos:2014}.
Simultaneous nighttime near infrared (1.9\,$\mu$m) measurements on 
Cerro Chajnantor and
183\,GHz water vapor line measurements at APEX
in 2009--2011 
indicated a PWV ratio of 0.5--0.6 \citep{konishi:2015}.

\section{ Summary}

Identical tipping radiometers were deployed to Maunakea, 
to the South Pole,
and to the environs of Cerro Chajnantor in 
northern Chile to measure the broadband 350\,$\mu$m and 200\,$\mu$m 
atmospheric transparency.
These measurements confirm all sites 
experience periods of excellent observing conditions.

The Chajnantor vicinity and the South Pole experience superior transparency
more often than Maunakea.
On the Chajnantor plateau and at the South Pole,
the best conditions are similar and  occur
roughly the same amount of the time (25\%).
At all sites, the transparency is better at night and during the winter.
Seasonal and interannual variations are similar in magnitude on Maunakea
whereas the seasonal variation is more pronounced at Chajnantor. 
The South Pole displays remarkable consistency from 
year-to-year.

Cerro Chajnantor enjoys significantly better transparency
than the Chajnantor plateau.
Most (75\%) of the time, the 350\,$\mu$m optical depth on Cerro Chajnantor 
is about 70\% of the optical depth on the plateau,
indicating  the typical water vapor scale height is 1100--1300\,m.
When conditions are good in the Chajnantor vicinity,
the 200\,$\mu$m optical depth is 3.2 times 
the 350\,$\mu$m optical depth. 

Because of seasonal and interannual  variations in
observing conditions and limitations on sky coverage
at any one place, no location alone is adequate
for all purposes.
Rather it is fruitful to have telescope facilities 
distributed across several sites.

\acknowledgments

Many thanks to the people who contributed to this project over the years.
E. Schartman constructed the instruments and initially deployed them;
R. Freund designed the electronics;
the South Pole winterover scientists
kept the instrument running;
S. Baca, E. Bufil, R. Chamberlin, A. Guyer, P. Nelson, 
K. Aird, E. Leitch, D. Marrone, 
R. Bustos, J. Cortes, C. Jara, F. Mu\~noz,
G. Gull, C. Henderson,
A. Ot\'arola, R. Reeves, R. Rivera, and G. Valladares,
provided invaluable help with  deployments;
S. Paine kindly measured the window transparencies;
P. Ade provided some of the filters;
M. Holdaway provided advice on interpretation;
E. Young allowed us to use his FTS;
K. Xiao did the added window experiments at
the South Pole;
and the referee offered several constructive comments.
APEX, AST/RO, CSO, CBI, QUIET, JCMT, NRAO/ALMA, SMA, and SPT provided space, 
power, and network connections.
APEX provides open access to their meteorological data.
Access to Cerro Chajnantor was possible because the University of
Tokyo constructed a road.

Development of the instruments was supported by Carnegie Mellon
University and the National Radio Astronomy Observatory.
Deployment to the South Pole was supported by the
Center for Astrophysical Research in Antarctica,
a National Science Foundation Science and Technology Center
operated under cooperative agreement.
The National Radio Astronomy Observatory is a facility of 
the National Science Foundation operated under cooperative agreement
by Associated Universities, Inc.
The Caltech Submillimeter Observatory (CSO) was operated by the 
California Institute of Technology 
with support from the National Science Foundation
(AST-0838261).
CCAT site evaluation was carried out in the Parque Astron\'omico Atacama 
in northern Chile under the auspices of the Programa de Astronom\'ia, 
a program of the Comisi\'on Nacional de Investigaci\'on Cient\'ifica y 
Tecnol\'ogica de Chile (CONICYT).
CCAT site evaluation received partial support from the
National Science Foundation (AST-0431503).

\bibliographystyle{apj}
\bibliography{ms}

\end{document}